\documentclass[11pt]{article}
\usepackage{mathrsfs, amsmath, amssymb,cite,array}
\usepackage{epsfig,graphicx,cases,subfig,url,fullpage}
\usepackage{color,array,txfonts,MnSymbol,float}

\title{Core-Periphery Structure in Networks}

\author{
M. Puck Rombach\footnotemark[2]
\and
Mason A. Porter\footnotemark[3]
\and
James H. Fowler\footnotemark[4]
\and
Peter J. Mucha\footnotemark[5]
}

\date{\today}

\begin{document}\maketitle
\renewcommand{\thefootnote}{\fnsymbol{footnote}}
\footnotetext[2]{Oxford Centre for Industrial and Applied Mathematics, Mathematical Institute, University of Oxford, {\tt rombach@maths.ox.ac.uk}}
\footnotetext[3]{Oxford Centre for Industrial and Applied Mathematics, Mathematical Institute and
CABDyN Complexity Centre, University of Oxford, {\tt porterm@maths.ox.ac.uk}}
\footnotetext[4]{Department of Political Science and School of Medicine, University of California, {\tt jhfowler@ucsd.edu}}
\footnotetext[5]{Department of Mathematics
and Institute for Advanced Materials, Nanoscience \&
Technology, University of North Carolina, {\tt mucha@unc.edu}}

\renewcommand{\thefootnote}{\arabic{footnote}}

\begin{abstract} Intermediate-scale (or `meso-scale') structures in networks have received considerable attention, as the algorithmic detection of such structures makes it possible to discover network features that are not apparent either at the local scale of nodes and edges or at the global scale of summary statistics.  Numerous types of meso-scale structures can occur in networks, but investigations of such features have focused predominantly on the identification and study of community structure.  In this paper, we develop a new method to investigate the meso-scale feature known as \emph{core-periphery structure}, which entails identifying densely-connected core nodes and sparsely-connected periphery nodes.  In contrast to communities, the nodes in a core are also reasonably well-connected to those in the periphery.  Our new method of computing core-periphery structure can identify multiple cores in a network and takes different possible cores into account. We illustrate the differences between our method and several existing methods for identifying which nodes belong to a core, and we use our technique to examine core-periphery structure in examples of friendship, collaboration, transportation, and voting networks.

\end{abstract}

\section{Introduction}

Networks are used to model systems in which entities, represented by nodes, interact with each other.  When representing a network as a graph, all of the connections are pairwise and hence represented by ties known as edges \cite{New10,Bocc06}.  Such a representation has led to numerous insights in the social, natural, and information sciences, and the study of networks has in turn borrowed ideas from all of these areas~\cite{Bara05}.

Networks can be described using a mixture of local, global, and intermediate-scale (meso-scale) perspectives.  Accordingly, one of the key uses of network theory is the identification of summary statistics for large networks in order to develop a framework to analyze and compare complex structures~\cite{New10}.  In such efforts, the algorithmic identification of meso-scale network structures makes it possible to discover features that might not be apparent either at the local level of nodes and edges or at the global level of summary statistics.

In particular, considerable effort has gone into algorithmic identification and investigation of a particular type of meso-scale structure known as community structure \cite{Port09,Fort09}, in which cohesive groups called `communities' consist of nodes that are connected densely to each other and the connections between nodes in different communities are comparatively sparse.  Myriad methods have been developed to detect network communities \cite{Port09,Fort09,Girv02,Girv04}, and this includes several that allow communities to overlap with each other \cite{Pall05,ahn10,Ball11}.  These efforts have led to insights in applications such as committee \cite{Port05} and voting \cite{Much10} networks in political science, friendship networks at universities \cite{traud} and other schools \cite{marta07}, protein-protein interaction networks \cite{anna}, and mobile telephone networks \cite{jp07}.

Although (and arguably because) studies of community structure have been very successful~\cite{Port09,Fort09}, the investigation of other types of meso-scale structures---often in the form of different `block models' \cite{doreian,Fort09}---have received much less attention than they deserve.  The type of meso-scale network structure that we consider in the present paper is known as \emph{core-periphery structure}.  The qualitative notion that social networks can have such a structure makes intuitive sense and has a long history in subjects like sociology \cite{laumann1976,doreian1985}, international relations \cite{wallerstein1974,steiber1979,chase1989,smithwhite}, and economics \cite{krugman1996}. The most popular quantitative method to investigate core-periphery structure was proposed by Borgatti and Everett in 1999 \cite{Borg99}.  Since then, various notions of core-periphery structure have been developed \cite{Holme05,Silva08,knotty-cent,jure2013,randomwalk2013}, but most examinations of core-periphery structure  still rely on implementations of the methods in Ref.~\cite{Borg99} or ~\cite{Comr62} in the software package UCInet \cite{ucinet}.

By computing a network's core-periphery structure, one attempts to determine which nodes are part of a densely connected core and which are part of a sparsely connected periphery. Core nodes should also be reasonably well-connected to peripheral nodes, but the latter are not well-connected to a core or to each other. Hence, a node belongs to a core if and only if it is well-connected both to other core nodes and to peripheral nodes. A core structure in a network is thus not merely densely connected but also tends to be `central' to the network (e.g., in terms of short paths through the network).  The goal of quantifying various notions of `centrality', which are intended to measure the importance of a node or other network component \cite{faust,New10}, also helps to distinguish core-periphery structure from community structure.  Additionally, networks can have nested core-periphery structure as well as both core-periphery structure and community structure \cite{Lesk08,jure2013}, so it is desirable to develop algorithms that allow one to simultaneously examine both types of meso-scale structure. 

\begin{figure}[ht]
\begin{center}
\begin{minipage}[c]{0.19\linewidth}
\centering
\includegraphics[scale=.2]{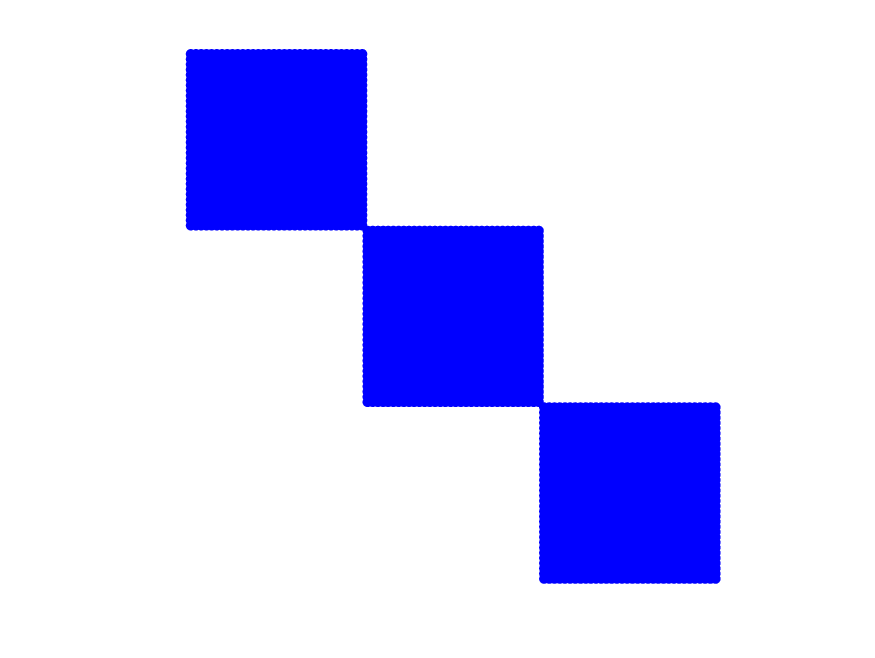}\\(a)
\end{minipage} \hspace*{3pt}
\begin{minipage}[c]{0.2\linewidth}
\centering
\includegraphics[scale=.2]{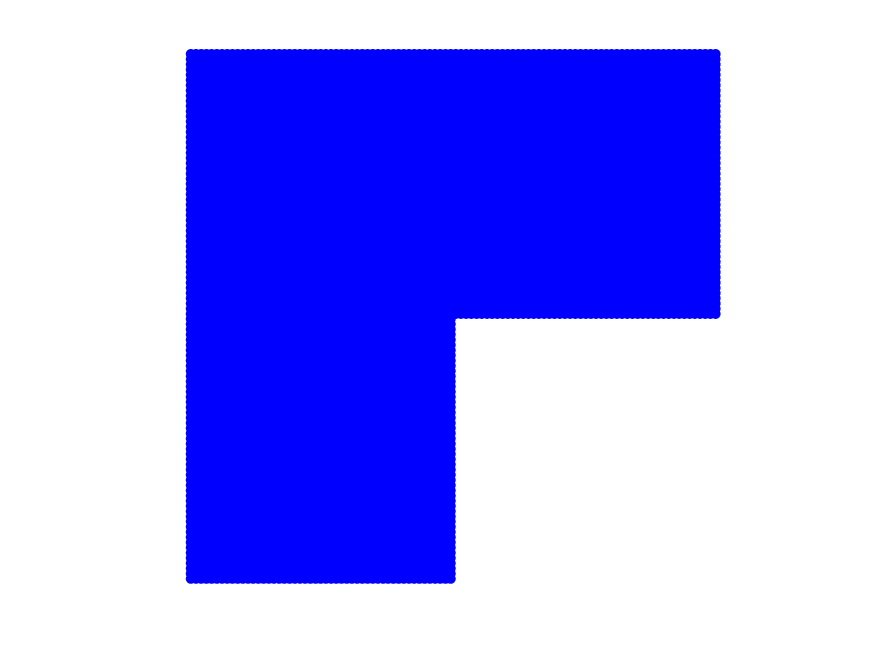}\\(b)
\end{minipage} \hspace*{3pt}
\begin{minipage}[c]{0.2\linewidth}
\centering
\includegraphics[scale=.2]{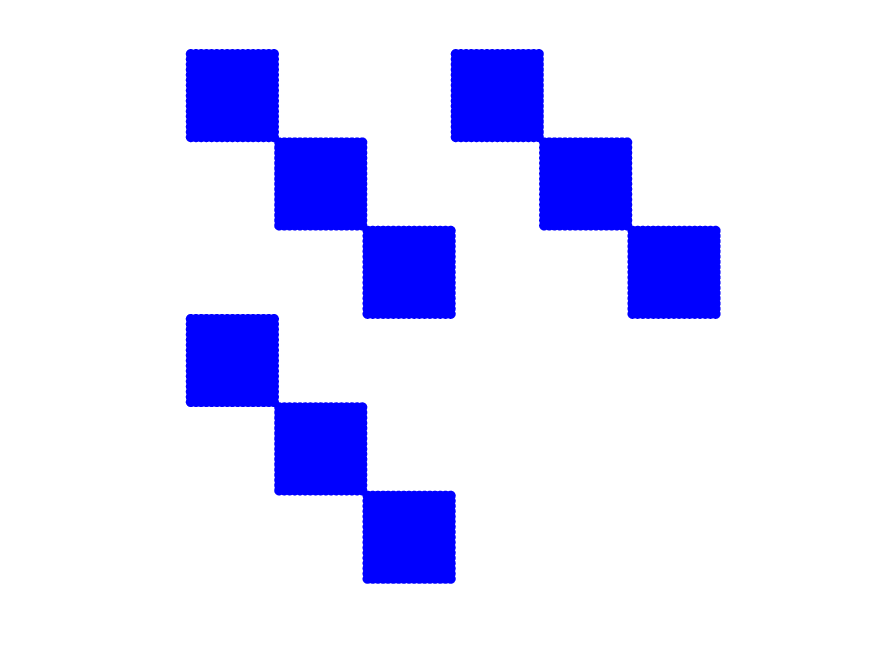}\\(c)
\end{minipage}\hspace*{3pt}
\begin{minipage}[c]{0.2\linewidth}
\centering
\includegraphics[scale=.17]{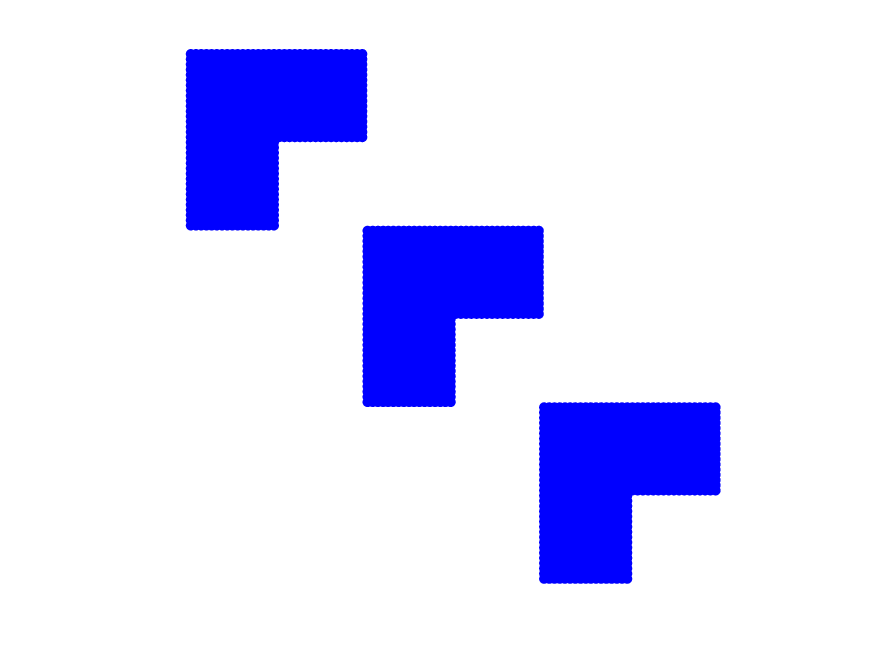}\\(d)
\end{minipage}
\end{center}
\caption{Examples of network block models. (a) Community structure, (b) core-periphery structure, (c) global core-periphery structure with local community structure, and (d) global community structure with local core-periphery structure. Note that (c) and (d) are equivalent.
}\label{comcorim}
\end{figure}

In Fig.~\ref{comcorim}, we show images of the adjacency matrices of idealized block models that illustrate (a) community structure, (b) core-periphery structure, (c) a global core-periphery structure with a local community structure, and (d) a global community structure with a local core-periphery structure.  By permuting rows and columns of the adjacency matrix, one can see that (c) and (d) are equivalent.

Several results underscore the importance of considering core-periphery structure in addition to community structure.  For example, Chung and Lu \cite{Chung02} showed that power-law random graphs, in which the number of nodes of degree $k$ is proportional to $k^{-\beta}$, almost surely contain a dense subgraph that has short distance to almost all other nodes in a graph when the exponent $\beta \in (2,3)$.  This suggests that it is sensible for networks with heavy-tailed degree distributions to contain some sort of cohesive core, and there is strong evidence that this is indeed the case in many real-world networks (such as many social networks and the World Wide Web) \cite{New10,Bocc06,faust}.  Moreover, core-periphery structure and community structure provide complementary lenses in which to view meso-scale network structures \cite{jure2013}.

Nodes of particularly high degree (which are sometimes called `hubs') occur in many real-world networks and can pose a problem for community detection, as they often are connected to nodes in many parts of a network and can thus have strong ties to several different communities.  For instance, such nodes might be assigned to different communities when applying different computational heuristics using the same notion of community structure \cite{good2010}, and it becomes crucial to consider their strengths of membership across different communities (e.g., by using a method that allows overlapping communities) \cite{Ball11,ahn10}.  In such situations, the usual notion of a community might not be ideal for achieving an optimal understanding of the meso-scale network structure that is actually present, and considering hubs to be part of a core in a core-periphery structure might be more appropriate \cite{Lesk08}.  For example, one can consider communities as tiles that overlap to produce a network's core \cite{jure2013}.

The rest of this paper is organized as follows. We first describe several previously proposed methods for detecting core-periphery structure in networks before presenting our new method, which computes a continuous value along a core-periphery spectrum and thereby yields a centrality measure based on core-periphery structure. We illustrate our method using a set of synthetic (computer-generated) benchmark random networks with a known core. We then apply our method to several real networks: the Zachary Karate Club, co-authorship networks of network scientists, a voting-similarity network of United States Senators, and the London Underground (`The Tube') transportation network. We conclude by summarizing our results, and we then present additional results and discussion in the Appendix.

\section{Detecting Core-Periphery Structure} \label{prevwork}

\subsection{Existing Methods}\label{exmeth}

Intuitively, one expects many real networks to possess some sort of core-periphery structure as part of their meso-scale structure. Perspectives proposed to examine core-periphery structure in a network include block models \cite{Borg99}, $k$-core organization \cite{Holme05}, consideration of connectivity information and short paths through a network \cite{Silva08,knotty-cent,randomwalk2013}, and overlapping of communities \cite{jure2013}.

The most popular notion of core-periphery structure in networks was developed by Borgatti and Everett \cite{Borg99}, who proposed algorithms for detecting both discrete and continuous versions of core-periphery structure in weighted, undirected graphs. Their discrete notion of core-periphery structure is based on comparing a network to a block model that consists of a fully-connected core and a periphery that has no internal edges but is fully connected to the core. Their method aims to find a vector $C$ of length $N$ whose entries can either be $1$ or $0$.  The $i^{\mathrm{th}}$ entry $C_i$ equals $1$ if the corresponding node is assigned to the core, and it equals $0$ if the corresponding node is assigned to the periphery.  Let $C_{ij}=1$ if $C_i=1$ or $C_j=1$, and let $C_{ij}=0$ otherwise. Define
\begin{equation}
	\rho_C = \sum_{i,j} A_{ij} C_{ij}\,,
\end{equation}
where the adjacency-matrix element $A_{ij}$ represents the weight of the tie between nodes $i$ and $j$, and it equals $0$ if nodes $i$ and $j$ are not connected. This method to compute a discrete core-periphery structure seeks a value of $\rho_C$ that is high compared to the expected value of $\rho_C$ if $C$ is shuffled such that the number of $1$ and $0$ entries are preserved but their order is randomized.  The output is the vector $C$ that gives the highest $z$-score for $\rho_C$.

As a variant discrete notion of core-periphery structure, Borgatti and Everett defined \cite{Borg99}
\begin{equation}
	C_{ij}=\begin{cases}
		1\,, &\text{if~}C_i \text{~and~} C_j=1  \,,\\
		a \in [0,1] \,, &\text{if~}C_i=1\text{~xor~}C_j=1  \,,\\
		0\,, &\text{otherwise}\,,
	\end{cases}
\end{equation}

where `xor' denotes an `exclusive or' operation. Borgatti and Everett also defined a continuous notion of core-periphery structure in which a node is assigned a `coreness' value of $C_i$ and $C_{ij} = C_i \times C_j = a$. Our method to study core-periphery structure in weighted, undirected networks (see Section \ref{ours}) is motivated by this continuous formulation of Borgatti and Everett. In UCInet \cite{ucinet}, the suggested heuristic for computing continuous core-periphery scores is the MINRES method \cite{Comr62,Boyd10}. MINRES seeks a vector $C$ such that the adjacency matrix is approximated by $CC^T$. The approximation minimizes
the off-diagonal sums of squared differences. It thus seeks to find a $C$ that minimizes $\sum_i\sum_{j \neq i}[A_{ij}-C_iC_j]^2$. Taking a partial derivative with respect to each element of $C$ gives 
\begin{equation}\label{minres}
	C_i=\frac{\sum_{j \neq i} A_{ij}C_j}{\sum_{j \neq i} C_j^2}\,, 
\end{equation}
which in turn yields an iterative process for computing the MINRES vector. Observe that this vector will in many cases be similar to the leading eigenvector of the adjacency matrix.

Holme defined a core-periphery coefficient \cite{Holme05}
\begin{equation}
	c_{cp}(G)=\frac{C_C\left(V_{\mathrm{core}}(G)\right)}{C_C\left(V(G)\right)}-\left\langle \frac{C_C\left(V_{\mathrm{core}}(G')\right)}{C_C\left(V(G')\right)} \right\rangle_{G' \in \mathcal{G}(G)}\,,
\end{equation}
where $V$ is the set of nodes of an unweighted and undirected graph $G$, the angled brackets indicate averaging, and $\mathcal{G}(G)$ is an ensemble of graphs with the same degree sequence as $G$.  Additionally,
\begin{equation}
	C_C(U)=\left( \left\langle \left\langle P(i,j) \right\rangle _{j \in V \backslash \lbrace i \rbrace } \right\rangle _{i \in U} \right)^{-1}\,,
\end{equation}
and $P(i,j)$ is the distance (i.e., number of edges in the shortest path) between nodes $i$ and $j$.  A $k$-core of the graph $G$ is a maximal connected subgraph in which all nodes have degree at least $k$, and $V_{\mathrm{core}}$ is the $k$-core with maximal $C_C(U)$. Using $k$-cores to examine core-periphery structure is computationally fast (and we note that one could, in principle, generalize Holme's method for weighted graphs using a notion of a weighted $k$-core \cite{havlin-weighted}), but it entails extremely strong restrictions on the notion of a network core. Philosophically, we view it as analogous to requiring a network community to be a clique.

One expects a core of a network to have high connectivity to other parts of the network, so Da Silva \emph{et al}.\ introduced a measure of connectivity known as network \emph{capacity} \cite{Silva08}:
\begin{equation}
	K=\sum_{l=1}^M P_l^{-1}\,,
\end{equation}
where $M$ is the total number of connected pairs of nodes and $P_l$ is the length of the shortest path between the $l^{\mathrm{th}}$ pair of nodes.  Da Silva \emph{et al}.\ then defined a core coefficient as $cc=N'/N$, where $N$ is the total number of nodes in the network, $N'$ satisfies $\sum_{m=0}^{N'}K_m=0.9 \sum_{u=0}^{N}K_u$, and $K_m$ is the capacity of the network after the removal of $m$ nodes. (One could define a more general notion using a parameter instead of the specific value 0.9.) The nodes are removed in order of closeness centrality, which is defined as the mean shortest path from a node to each of the other nodes in a network~\cite{Silva08}. Note that in the remainder of this paper, we will use the following definition for the closeness centrality of a node $j$ (there are several different definitions available in the literature \cite{New10}):
\begin{equation*}
	CC_j=\frac{1}{N}\sum_{i \in V} P(i,j)\,,
\end{equation*}
where $P(i,j)$ is the sum of edge weights in a shortest path in the context of weighted networks.  Da Silva \emph{et al}.\ considered only binary networks, but their method can be generalized straightforwardly to weighted networks.

Other recent ideas for examining core nodes in a network include the computation of `knotty centrality' \cite{knotty-cent} (which attempts to discover nodes that have high geodesic betweenness centrality but which need not have high degree), the identification of cores based on collections of nodes in overlapping communities \cite{jure2013}, and the use of random walkers \cite{randomwalk2013}.

\subsection{Our Method}\label{ours}

Our method to study core-periphery structure in weighted, undirected networks is motivated by the continuous formulation of Borgatti and Everett \cite{Borg99} that we described above. However, our method takes cores of different size and shapes into account. It thereby gives credit to all nodes that take part in a core, and it weights this credit by the quality of the associated core.  As we discuss below, we employ a \emph{transition function} to interpolate between core and periphery nodes. Additionally, we construct elements $C_{ij}$ of a \emph{core matrix} to compute the quality of a core. We will present several viable choices for both the transition function and the core matrix.

We define the \emph{core quality}
\begin{equation}\label{R-general}
	R_{\gamma} = \sum_{i,j}A_{ij} C_{ij}\,,
\end{equation}
where $\gamma$ is a vector that parametrizes the core quality (see the discussion below), the elements $C_{ij}$ of the core matrix are given by $C_{ij} = f(C_i,C_j)$, and $C_i \geq 0$ is the \emph{local core value} of the $i^{\mathrm{th}}$ node.  The local core values are elements of a \emph{core vector} $C$. Our example calculations in this paper usually use a product form
\begin{equation}\label{product}
	C_{ij}=C_i C_j\,,
\end{equation}
but we discuss other viable choices in Section \ref{corematrix}.

We seek a core vector $C$ that maximizes $R_{\gamma}$ and is a normalized (so that its entries sum to $1$) shuffle of the vector $C^*$ whose components $C^*_i = g(i)$ are determined using a \emph{transition function} $g$.  The number of components of the vector $C^*$ is equal to the number of nodes in the network, and $C^*_i$ gives the local core value of the $i^{\mathrm{th}}$ node. Our example calculations in this paper usually use the transition function given by the sharp (because it has a discontinuous derivative) function 
\begin{align}\label{trans1} 
	C^*_{i}(\alpha,\beta) = g_{\alpha,\beta}(i) = \begin{cases}
		\frac{i(1-\alpha)}{2\beta}\,,\qquad\qquad &i \in \{1, \ldots, \beta N\}\,, \\
		\frac{(i-\beta)(1-\alpha)}{2(N-\beta)}+\frac{1+\alpha}{2}\,, \qquad\qquad &i \in \{\beta N + 1, \ldots, N\}\,.
		\end{cases}
\end{align}
{The parameter $\beta$ sets the size of the core: as $\beta$ varies from $0$ to $1$, the number of nodes included in the core varies from $N$ to $0$. The parameter $\alpha$ sets the size of the score jump between the highest scoring periphery node and the lowest scoring core node. In the limit in which $\alpha=1$, this yields a discrete classification (discontinuous function) into a unique core and unique periphery that assigns each node to either the core or the periphery. 

With the transition function (\ref{trans1}) and the product form (\ref{product}) for the core-matrix elements, the core quality is given by
\begin{equation}\label{R}
	R_\gamma = R_{\alpha,\beta} = \sum_{i,j}A_{ij} C_{ij} = \sum_{i,j}A_{ij}C_i C_j\,.
\end{equation}
For a given value of $\gamma = (\alpha,\beta)$, we seek a shuffle $C$ of $C^*$ such that $R_\gamma$ is maximized. 

For any choice of core matrix and transition function, we define the aggregate \emph{core score} of each node $i$ as
\begin{equation}\label{totalcorescore}
	CS(i)=Z\sum_{\gamma}C_i(\gamma) \times R_{\gamma}\,,
\end{equation}
where the normalization factor $Z$ is chosen so that $\max_k[CS(k)] = 1$, where $k \in \{1,\ldots, N\}$ indexes the nodes. A core score gives a notion of network centrality \cite{faust,New10}.  As discussed above, our usual choice in this paper is maximize the core quality (\ref{R}) that uses the product form (\ref{product}) for the core matrix and the sharp transition function (\ref{trans1}) to interpolate between core and periphery nodes.  See Section \ref{corematrix} for a discussion of other choices for constructing the core matrix and Section \ref{transition} for other choices of transition function.

In the results that we present in this paper, we assign the values of $C^*_i(\alpha,\beta)$ to the nodes to obtain a $C_i(\alpha,\beta)$ that maximizes $R_{\alpha,\beta}$ using a simulated-annealing algorithm \cite{Kirk83}. (See the Appendix for details of the procedure.) Other computational heuristics can, of course, be faster.  In all of our examples using a two-parameter transition function, we sample $\alpha$ and $\beta$ uniformly over a discretization of the square $[0,1] \times [0,1]$.  In particular, we always use $\alpha  = \beta = [0.01:0.01:1]$ (in {\sc Matlab} notation).
It is also interesting to consider the core quality of specific values of $\alpha$ and $\beta$, and one could in principle improve the speed of our general approach by developing procedures for choosing $\alpha$ and $\beta$ selectively in a manner that takes advantage of the structure of particular networks or families of networks.  Indeed, the a priori choice of which values of $\alpha$ and $\beta$ to sample is a difficult but interesting question.  The purpose of this paper is to introduce a novel notion of core-periphery structure and to demonstrate why it is interesting using a variety of examples, so we leave the aforementioned issues for future consideration.

\subsubsection{Functional Forms for Elements of Core Matrix}\label{corematrix}

In most of the calculations in this paper, we construct the core-matrix elements $C_{ij}$ using a product form $C_{ij} = C_i C_j$.  However, other choices are also viable.

An idealized core-periphery structure entails that core nodes are well-connected to other core nodes as well as to periphery nodes and that periphery nodes are not well-connected to each other. Let $v_1$ and $v_2$ be core nodes and let $w_1$  and $w_2$ be peripheral nodes. We then want $C_{w_1w_2}$ to be small and $C_{v_1v_2}$ and $C_{v_iw_j}$ to be large.  For example, the block structure in panel (b) of Fig.~\ref{comcorim} satisfies these conditions.

As one can see from Fig.~\ref{Cfunction}, one can try to approximate such an idealized block structure using various ways of constructing $C_{ij}$.  For example, in addition to the product form (\ref{product}), one can instead use a $p$-norm and write
\begin{equation}\label{pnorm}
	C_{ij}= \Vert (C_i,C_j) \Vert_p =\sqrt[p]{ C_i^p+C_j^p}\,.
\end{equation}
As one considers progressively larger $p$, this will look more and more like an ideal core-periphery block model (in which core-core edges and core-periphery edges produce a value of 1 in a network adjacency matrix, but periphery-periphery edges produce a value of 0).

\begin{figure}
\begin{center}
\begin{minipage}[c]{0.3\linewidth}
\centering
\includegraphics[scale=.3]{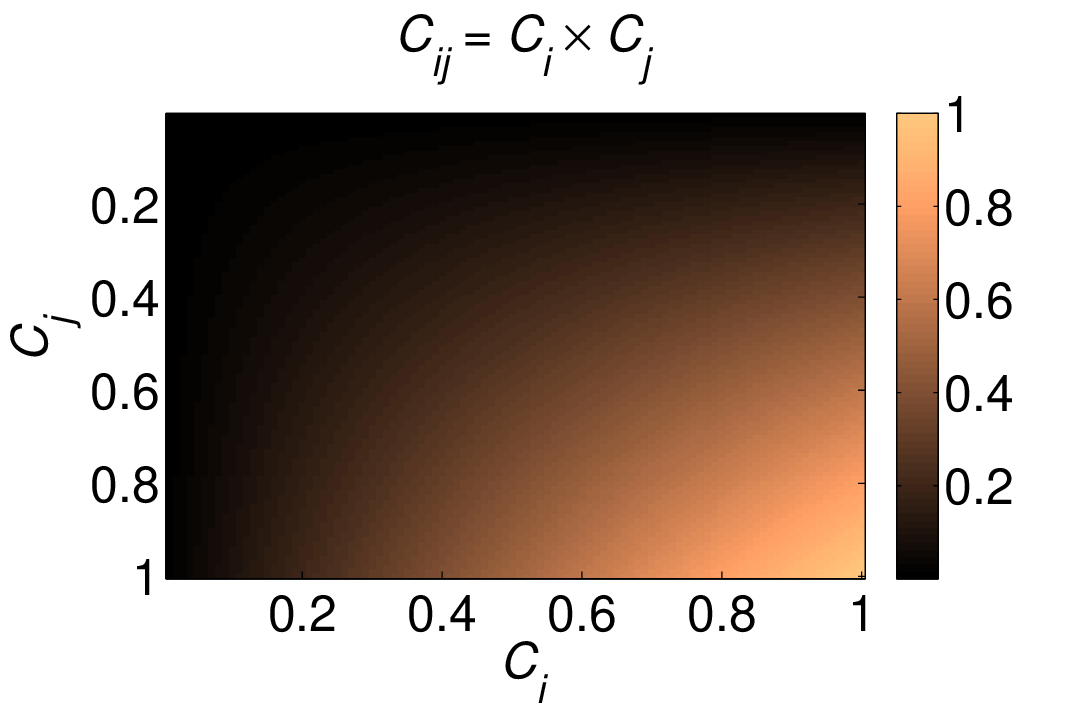}\\(a)
\end{minipage} \hspace*{3pt}
\begin{minipage}[c]{0.3\linewidth}
\centering
\includegraphics[scale=.3]{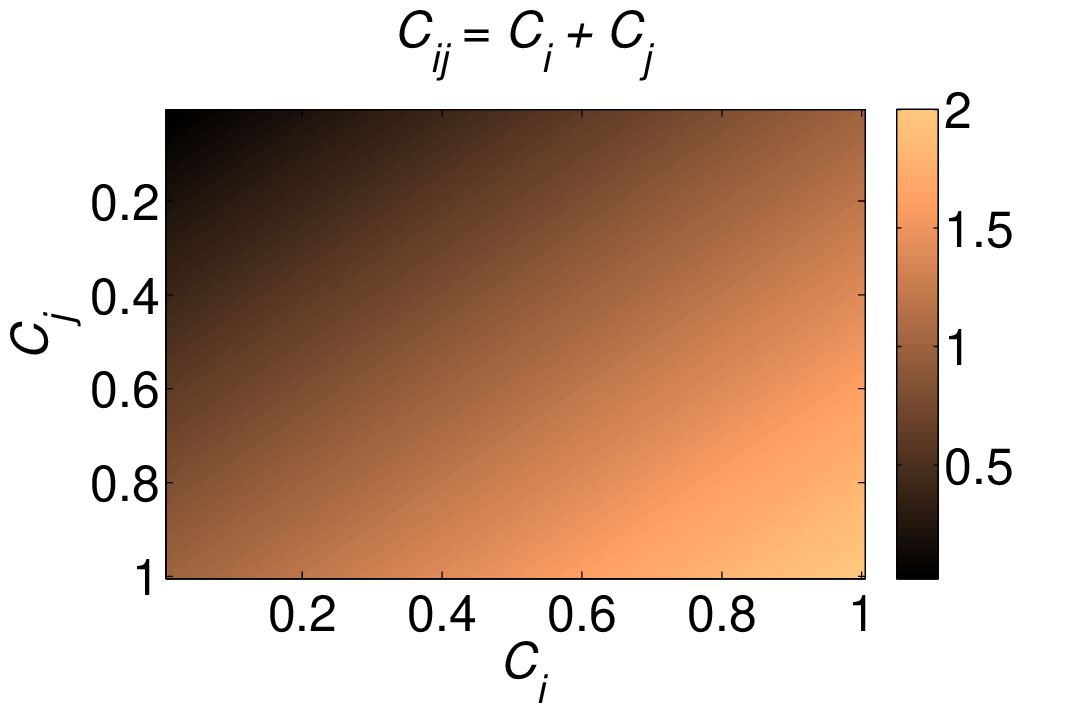}\\(b)
\end{minipage}\hspace*{3pt}
\begin{minipage}[c]{0.3\linewidth}
\centering
\includegraphics[scale=.3]{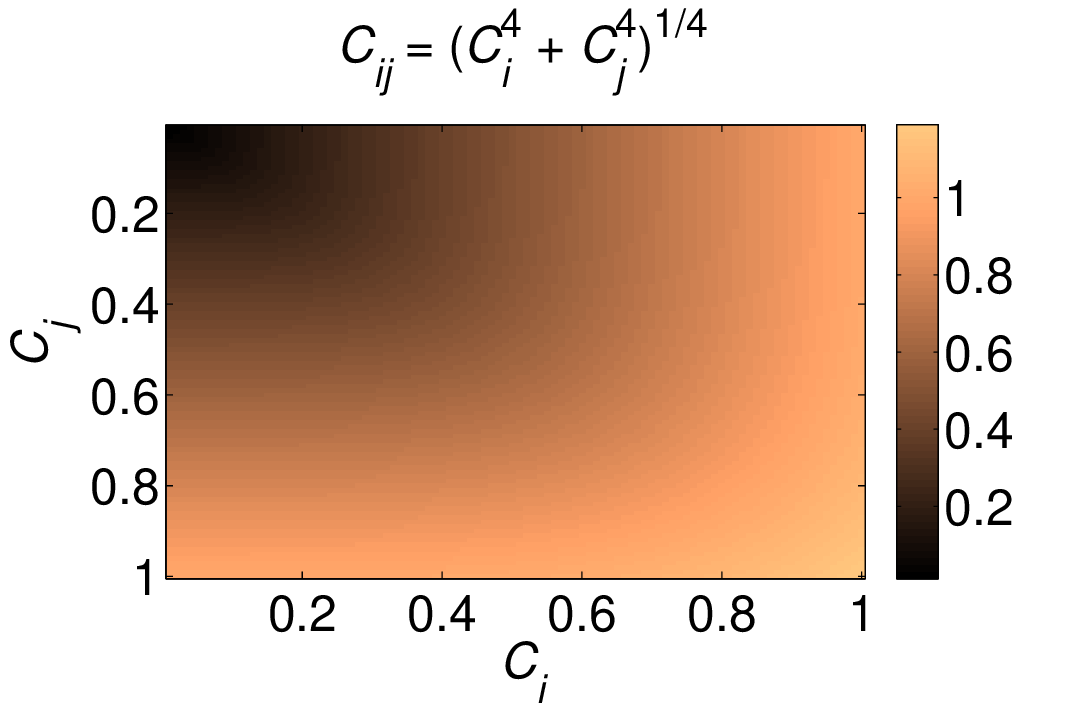}\\(c)
\end{minipage}
\end{center}
\caption{Several options for the core-matrix element $C_{ij}$ include (a) the product form $C_{ij}=C_i \times C_j$, (b) the 1-norm $C_{ij}= \Vert (C_i,C_j) \Vert_1=C_i + C_j$, and (c) the 4-norm $C_{ij}= \Vert (C_i,C_j) \Vert_4=\sqrt[4]{C_i^4 + C_j^4}$.
}\label{Cfunction}
\end{figure}

\subsubsection{Transition Function}\label{transition}

Our methodology to compute core-periphery structure entails choosing a transition function to interpolate between core and periphery nodes. In most of the calculations in this paper, we use the sharp two-parameter function (\ref{trans1}) to illustrate our approach.  However, there are many other viable choices for the transition function.

One variant is to construct the vector $C^*$ using a smooth transition function $g(i)$.  For example, one possibility is
\begin{equation}\label{trans2}
	C^*_i(\alpha,\beta) = g_{\alpha,\beta}(i) = \frac{1}{1+\exp\left\{-(i-N\beta) \times \tan(\pi \alpha /2)\right\}}\,,
\end{equation}
which has parameters $\alpha \in [0,1]$ and $\beta \in [0,1]$. The parameter $\alpha$ sets the sharpness of the boundary between the core and the periphery. The value $\alpha = 0$ yields the fuzziest boundary and $\alpha = 1$ gives the sharpest transition: as $\alpha$ varies from $0$ to $1$, the maximum slope of $C^*$ varies from 0 to $\infty$.  The parameter $\beta$ sets the size of the core: as $\beta$ varies from $0$ to $1$, the number of nodes included in the core varies from $N$ to $0$. 

Another option, which allows our method to be significantly faster, is a transition function that has only one parameter.  One can choose such a parameter to control the size of the core, the sharpness of the boundary, or some combination of the two.  For example, one possibility is
\begin{equation}\label{trans3}
	 C_i^*(\alpha) = g_\alpha(i) = \frac{1}{2}\mathrm{tanh}\left(8\exp\left\{\frac{-10*(\alpha-N/2)^2}{2} \right\}(i-\alpha)+1\right)\,.
\end{equation}
We plot (\ref{trans3}) for various values of $\alpha$ in Fig.~\ref{tanh}.  One can then average over values of $\alpha$ to produce aggregate core scores.

\begin{center}
\begin{figure}[!htbp]
\begin{minipage}[c]{\linewidth}
\begin{center}
\includegraphics[scale=.65]{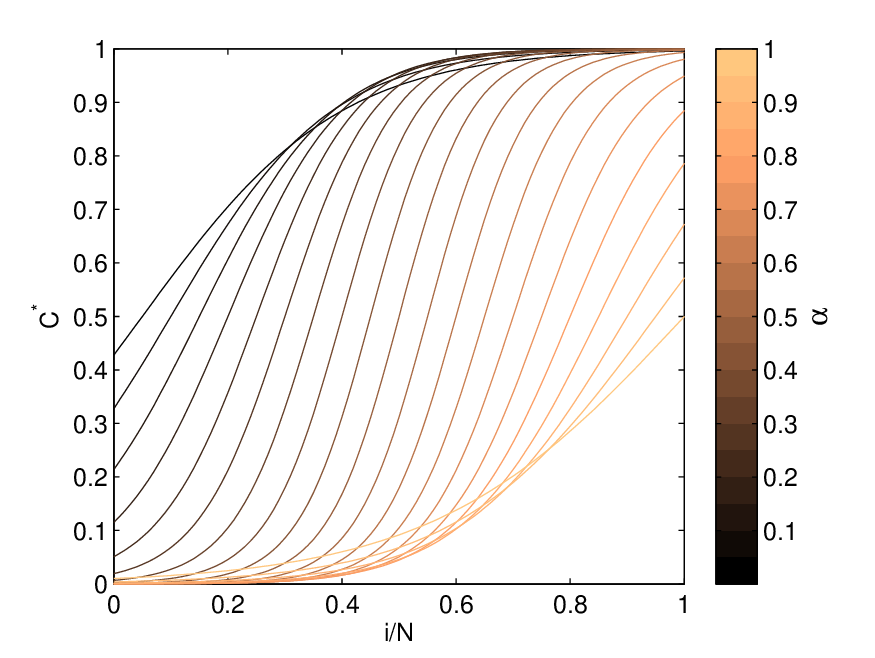}
\end{center}
\end{minipage}
\caption{An example of a one-parameter transition function in which the parameter $\alpha$ controls both the size of a network core and the sharpness of the boundary between core and periphery nodes.
}\label{tanh}
\end{figure}
\end{center}

In this paper, we calculate aggregate core scores using formulations with both two-parameter and one-parameter transition functions.  In the former case, we always average over 10000 values of $(\alpha,\beta)$ that are sampled uniformly from $[0,1] \times [0,1]$.  (In particular, we use $\alpha  = \beta = [0.01:0.01:1]$.)  In the latter case, we always average over 10000 values of $\alpha$ that are uniformly sampled from $[0,1]$. (In particular, we use $\alpha = [0.0001:0.0001:1]$.)

\subsubsection{Interpreting Core Scores}\label{interpret}

There are several ways to use and interpret the results of our approach for studying core-periphery structure.  One can average over a set of parameter values---e.g., in the $(\alpha,\beta)$ parameter plane if one uses a two-parameter transition function---and obtain a set of aggregate core scores that yield a continuous centrality measure for the networks in a network.  Alternatively, one can use the core-periphery structure at a single set of parameter values, such as the one that produces the largest value of the core quality $R$ (\ref{R-general}). (See the discussion of the Zachary Karate Club network in Section \ref{karate}.)  Sometimes, as with the London Underground network in Section \ref{tube}, one can observe a clear dichotomy between core and periphery nodes after calculating continuous core scores. Finally, it can be useful to impose a specific core size in advance (and thereby dichotomize core and periphery nodes), as we do with the synthetic benchmark networks in Section \ref{artcore}.

The flexibility described in the above paragraph is a beneficial feature of our method, which can be used either to produce a continuum of core scores or a discrete classification of core versus periphery.  The utility for both of these perspectives, and hence the desirability for the development of methods to study core-periphery structure that have such flexibility, was recognized more than two decades ago \cite{Borg99,smithwhite,chase1989}.  For example, studies of international relations include vehement arguments as to whether countries should be classified discretely (e.g., into core, semiperipheral, and peripheral countries) or along a continuum \cite{smithwhite}, and methods that can produce both discrete and continuous perspectives on core-periphery structure ought to be helpful for studying such applications.

\section{Synthetic Benchmark Networks}\label{artcoresec}

In this section, we examine our method using an ensemble of random networks with an imposed core-periphery structure to demonstrate that it performs well at detecting the kind of core-periphery structure envisioned by Borgatti and Everett \cite{Borg99}. We then consider lattice networks, which do not have any meaningful core-periphery structure.

\subsection{Imposed Core-Periphery Structure}\label{artcore}

We develop a family of synthetic networks that only have a core-periphery structure [see Fig.~\ref{comcorim}(b)], and we use $CP(N,d,p,k)$ to denote this ensemble of networks.  (We will consider networks with both core-periphery structure and community structure when we examine real networks.  For example, see the London Underground network in Section \ref{tube} and the network of network scientists in Section \ref{nns}.) Each network in the ensemble $CP(N,d,p,k)$ has $N$ nodes, where $dN$ of the nodes are core nodes, $(1-d)N$ of the nodes are peripheral nodes, and $d \in [0,1]$. The edges are assigned independently at random. The edge probabilities for periphery-periphery, core-periphery, and core-core pairs are $p$, $kp$, and $k^2p$, respectively.  Note that $p \in [0,1]$ and $ k \in [1,(1/p)^{1/2}]$.  We fix $N=100$, $d={1}/{2}$, and $p={1}/{4}$ and compute the core-periphery structure averaged over $100$ different instances of $CP(N,d,p,k)$ for each of the parameter values $k=1,1.1,1.2,\ldots,2$. In Fig.~\ref{artcore1}, we show our results of determining core nodes by computing the aggregate core score (\ref{totalcorescore}) with core quality (\ref{R}) and transition function (\ref{trans1}).The synthetic networks in $CP(N,d,p,k)$ possess a discrete core-periphery structure, whereas our method produces a continuous ranking, which we recall makes the aggregate core score a notion of centrality.

We also examine the results of attempting to determine the core nodes using various types of centrality \ref{exmeth}: closeness, degree, PageRank\cite{Page99}, geodesic node betweenness \cite{New10}, and MINRES \cite{Comr62}, which are designed to measure notions of node importance. We only test continuous node-ranking notions, which we evaluate by counting how many of the 50 core nodes---recall that the networks have $Nd = 50$ core nodes by construction---are placed in the top 50 according to each method.  (Alternatively, one can use information-theoretic diagnostics to evaluate the results of comparisons like this.) In Fig.~\ref{artcore}, we show the fraction of nodes that are correctly identified as one of the top 50 core nodes. When testing the methods, we used a random permutation of the labels of the nodes to prevent any bias. In this case, none of the tested methods should suffer from such a bias. (Note that our method starts the optimization with a random permutation of the vector $C^*$.)  We used our own implementation of MINRES for the calculation in this figure.

As we have indicated, our method examines core-periphery structure as a type of centrality.  Nodes are more likely to be part of a network's core if they have high strength (i.e., weighted degree) \emph{and} if they are connected to other core nodes.  Neither notion of importance is sufficient on its own.  Nodes with high degree are construed as important in many situations, and the latter idea is reminiscent of quantities like eigenvector centrality and PageRank centrality \cite{Page99}, which recursively define nodes as important based on having connections to other nodes that are important \cite{New10}.  We will also compare core scores with notions of centrality when we discuss political voting-similarity networks in Section \ref{voting}.

\begin{figure}[!h]
\begin{minipage}[c]{\linewidth}
\begin{center}
\includegraphics[scale=.65]{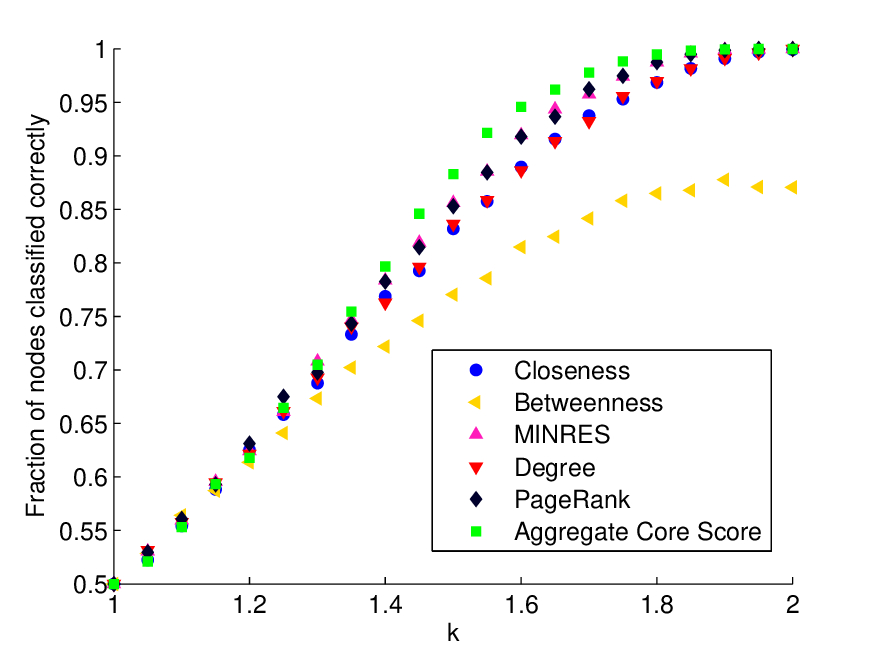}
\end{center}
\end{minipage}
\caption{Fraction of core nodes correctly identified by computing aggregate core score averaged over $100$ realizations of networks in the ensemble $CP(100,.5,.25,k)$. We compute the aggregate core score (\ref{totalcorescore}) using the core quality (\ref{R}) and the transition function (\ref{trans1}).
}\label{artcore1}
\end{figure}

\subsection{Lattices}\label{lattice}

As another example of a synthetic network, consider a lattice, which does not exhibit any meaningful core-periphery structure. (A lattice also does not have any meaningful community structure.) All nodes in a lattice have the same degree if one uses periodic boundary conditions. Moreover, lattices are \emph{symmetric}: for any two nodes, there exists a network automorphism that swaps the labelling of these two nodes. Thus, if one node is placed in the core and the other is placed in the periphery, then one could relabel the network in a way that would swap those assignments. Thus, for such networks, any assignment of core-periphery structure is arbitrary. The aggregate core score of \emph{every} node in a lattice converges to the same value (which is equal to 1) as one applies our method with increasingly high precision [i.e., using more values of $(\alpha ,\beta)$].

A possible concern about our methodology is that it might lead to false positives due to `forcing' different core-periphery structures on a network---especially given that we set the maximum aggregate core score to be 1, so every network will always have high scores.  However, as lattices illustrate, this does not necessarily lead to false positives. The aggregate score is an average over many computational runs (using different values of $\alpha$ and $\beta$), and symmetry guarantees that each node has an equal probability of being assigned a high score in a given run. Therefore, by taking averages over many runs, we see that the aggregate core score of each node is similar, and one converge to equal core scores in the limit of averaging over infinitely many runs.  Hence, our method correctly indicates that lattice networks have no meaningful core-periphery structure.

This example is simple, but it illustrates that one should examine not simply core-score magnitude but rather how core scores are distributed.  Just as with other centrality measures, this can be done visually, by computing the variance, or by computing a centralization \cite{faust}.

\section{Real Networks}\label{real}

In this section, we examine core-periphery structures in networks constructed using various real-world data sets.

\subsection{The Zachary Karate Club}\label{karate}

We first consider the infamous Zachary Karate Club network~\cite{Zach77}, which consists of friendship ties between 34 members of a university karate club in the United States in the 1970s. (In this paper, we use the unweighted version of this network.) A conflict led the club to split into two new clubs, and the (unweighted) Zachary Karate Club network has become one of the standard benchmark examples for investigations of community structure \cite{Port09,Fort09}. We visualize the network in Fig.~\ref{zach1}, where we have identified the nodes according to the split that occurred as a result of a longstanding disagreement between the instructor (Mr.~Hi) and the club president (John A.)\footnote{These names are pseudonyms introduced in Ref.~\cite{Zach77}}. These two primary actors are represented, respectively, by nodes 1 and 34.

In Table~\ref{zachtable}, we the show the nodes along with their aggregate core scores (\ref{totalcorescore}) computed using the core quality (\ref{R}) and the transition function (\ref{trans1}).  We also show the node degrees, which have a high positive correlation with the aggregate core scores. Unsurprisingly, the main actors (nodes 1 and 34) have the highest aggregate core scores. One can see additional structure by considering all values of the parameters $\alpha$ and $\beta$ rather than averaging over them.  (Recall that we consider $\alpha  = \beta = [0.01:0.01:1]$.) In particular, the fact that node $1$ has the highest aggregate core score does not imply that it has the highest value of $C_1^*(\alpha,\beta)$ for all $\alpha$ and $\beta$. In Fig.~\ref{zachcolour}, we show how the top node varies as a function of $\alpha$ and $\beta$.  Node $1$ has the highest core value only about $20\%$ of the time, whereas node $34$ is the top node about $74 \%$ of the time. However, the values for $\alpha$ and $\beta$ for which node 34 is the top node have lower core qualities $R$ (\ref{R}) on average than those for which node 1 is at the top.  Such nuances are invisible if one attempts to examine coreness using only the notion of degree. Figure ~\ref{zachcolour} also illustrates that we obtain different cores for different values of $\alpha$ and $\beta$.

Some of the nodes (e.g., 15, 16, 19, 21, and 23) in the Zachary Karate Club network are automorphs of each other (such nodes are \emph{role equivalent}\cite{roleequiv,doreian}), as one can swap their labels without changing the network structure.  In the limit as the number of runs in computing core-periphery structure becomes infinite, such nodes will be assigned the same aggregate core score.  See our discussion of lattice networks in Section \ref{lattice}.

\begin{center}
\begin{figure}[!h]
\begin{minipage}[c]{\linewidth}
\begin{center}
\includegraphics[scale=.8]{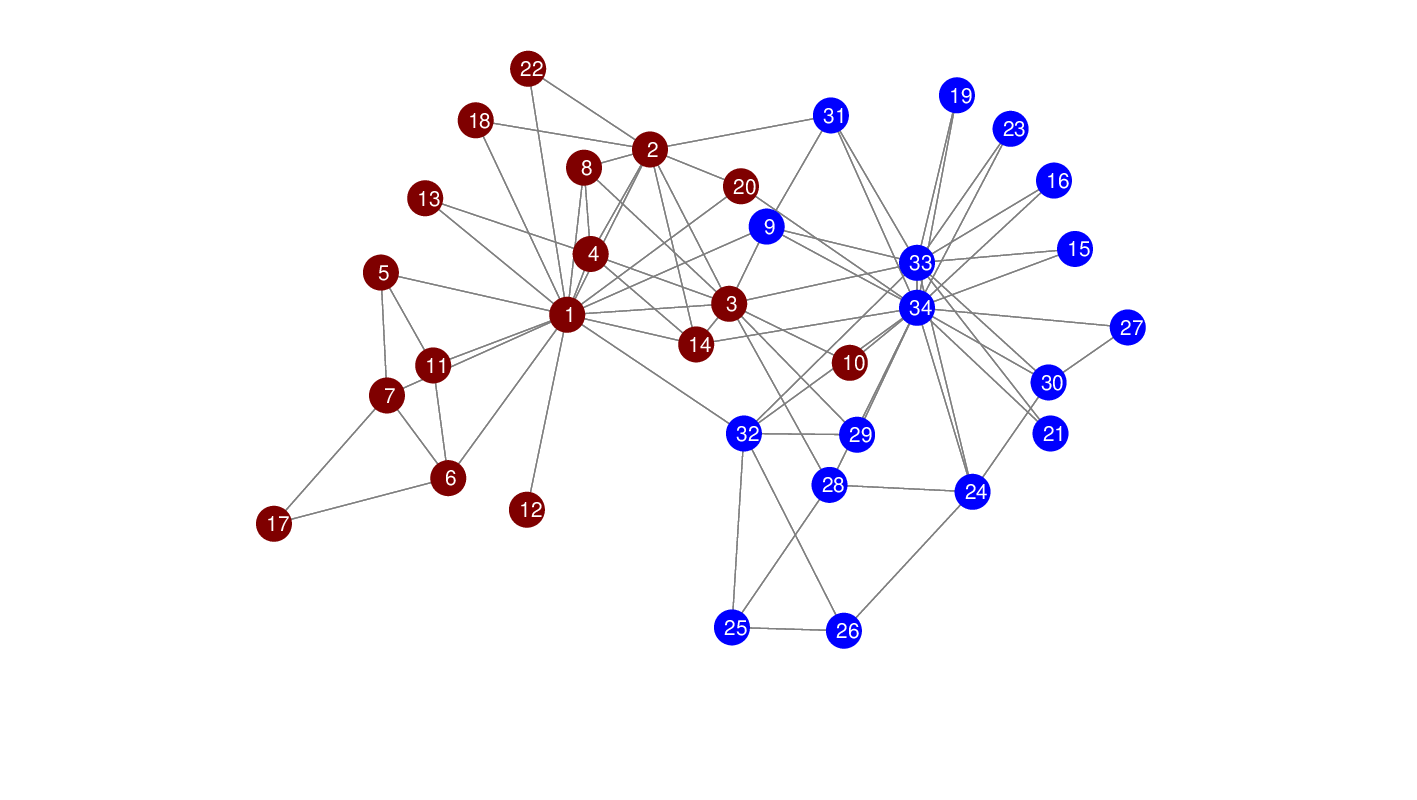}
\end{center}
\end{minipage}
\caption{The Zachary Karate Club network~\cite{Zach77}, which we visualize using the implementation of the Kamada-Kawai algorithm \cite{Kama88} in Ref.~\cite{Traud09}.  The colors represent the two groups into which the club split while it was under study.
}
\label{zach1}
\end{figure}
\end{center}

\begin{center}
\begin{table}
\caption{Nodes in the Zachary Karate Club network nodes along with their aggregate core scores (\ref{totalcorescore}) computed using the core quality (\ref{R}) and the transition function (\ref{trans1}).  We also give the node degrees}

\begin{minipage}[c]{\linewidth}
\begin{center}
{\footnotesize
\begin{tabular}{| l| l| l || l| l| l|}
\hline
 Node &  {Core Score} & Degree &  Node & {Core Score} & Degree \\
\hline
1 &1.0000 & 16 & 19 &.2255 & 2  \\

34 &.9951 & 17 & 15 &.2254 & 2\\

3 &.9702 & 10 & 21 &.2254 & 2\\

33 &.8719 & 12 & 23 &.2244  & 2 \\

2 &.8577 & 9 & 16 &.2244 & 2\\

9 &.7755 & 5 & 26 &.2196 & 3\\

14 &.7546 & 5&  25 &.2038 & 3\\

4 &.7537 & 6& 7 &.1840 & 4\\

8 &.6441 & 4 &6 &.1840 & 4\\

31 &.5849 &4 & 18 &.1787 & 2  \\

32 &.5377 & 6 &  22 &.1785 & 2\\

24 &.4661 & 5 & 11 &.1580 & 3  \\

20 &.4499 &3 & 5 &.1579 & 3\\

30 &.4152 & 4 & 13 &.1425 & 2\\

28 &.3957 & 4 & 27 &.1050 & 2 \\

29 &.3784 & 3& 12 &.0477 & 1 \\

10 &.2506 &2& 17 &.0343 & 2\\
   \hline
\end{tabular}\label{zachtable}}
\end{center}
\end{minipage}
\end{table}
\end{center}

\begin{center}
\begin{figure}[!h]
\begin{minipage}[c]{\linewidth}
\begin{center}
\includegraphics[scale=.6]{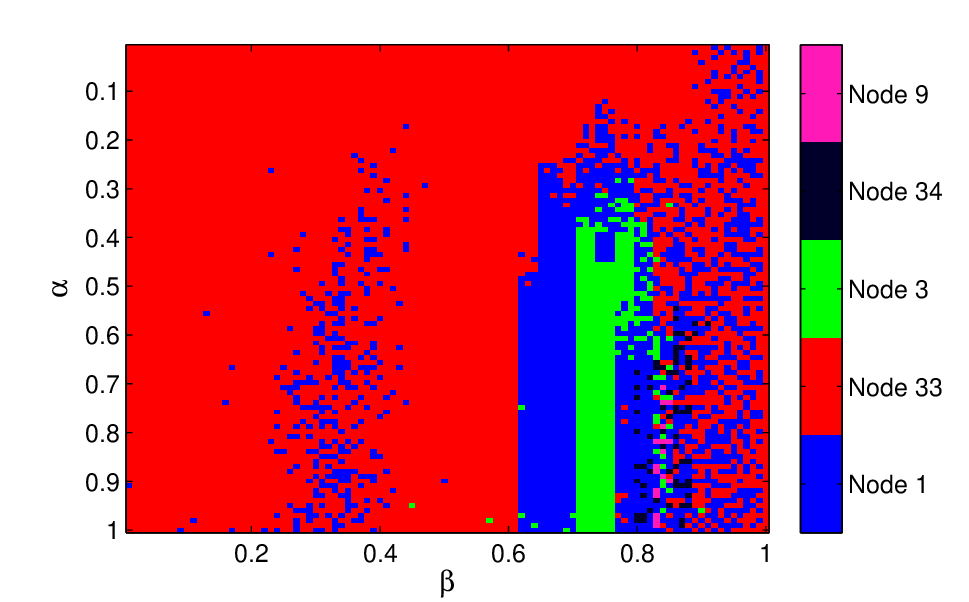}
\end{center}
\end{minipage}
\caption{The node of the Zachary Karate Club that has the top core score (i.e., $\mathrm{arg}\{\mathrm{max_k}(C)\}$, where $k \in \{1, \ldots , 34\}$ indexes the nodes) as a function of $\alpha$ and $\beta$.  We computed core scores using the core quality (\ref{R}) and the transition function (\ref{trans1}).
}\label{zachcolour}
\end{figure}
\end{center}

\begin{center}
\begin{figure}[!h]
\begin{minipage}[c]{\linewidth}
\begin{center}
\includegraphics[scale=.6]{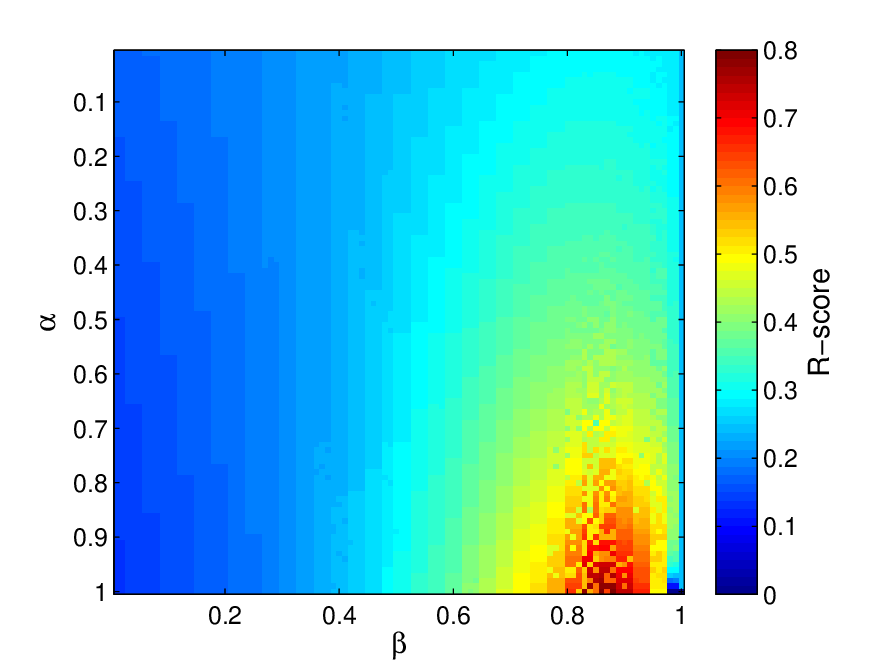}
\end{center}
\end{minipage}
\caption{Core quality $R$ (\ref{R}) of nodes in the Zachary Karate Club as a function of the parameters $\alpha$ and $\beta$.  We used the transition function (\ref{trans1}).
}\label{zachcontour}
\end{figure}
\end{center}

We illustrate this result by plotting the core quality $R$ \eqref{R} as a function of $\alpha$ and $\beta$ (see Fig.~\ref{zachcontour}).  The landscape of top core nodes can be complicated, especially as one considers larger networks, but examining it in a small network like the Zachary Karate Club is convenient for illustrating both how our method works and how it exposes multiple possible core-periphery structures in one network.  

\subsection{The London Underground}\label{tube}

One expects many metropolitan (metro) and subway transportation networks to exhibit a core-periphery structure \cite{marc-subway}.  To illustrate this, we compute core scores for the London Underground (`Tube') transportation network, which exhibits a strong core-periphery structure and a weak community structure.  We collected the data for this example using the website for the London Underground ({\tt http://www.tfl.gov.uk}).  The Tube network that we assembled has 317 nodes (one for each station) and weighted edges that represent the number of direct, contiguous connections between two stations. For example, Baker Street and Edgware Road share an edge of weight 2, as they are adjacent stations on both the Circle Line and the Hammersmith \& City Line. They are also connected by the Bakerloo Line; however, they are not adjacent stations on that line, so it does not affect the weight of the edge between them.

We partitioned the network into communities algorithmically by optimizing the modularity quality function \cite{New10,Port09,Fort09} using the Louvain \cite{Blond08} computational heuristic.  This splits the network into 21 communities, and the largest community that we obtained contains 19 nodes\footnote{The Louvain method is stochastic, so one can get slightly different network partitions in different runs of the algorithm.  We simply wanted a reasonable community structure as a means of comparison, so we used a single run of the algorithm in each situation for which we compute community structure.}.  Most of these communities consist of groups of stations on a single line.

In Table~\ref{tuberesult}, we show the results that we obtained for the London Tube network by computing aggregate core scores (\ref{totalcorescore}) using the core quality (\ref{R}) and the transition function (\ref{trans1}). We list the top ten stations and their corresponding aggregate core scores.

\begin{center}
\begin{table}[ht]
\begin{minipage}[c]{\linewidth}
\begin{center}
{\footnotesize
\begin{tabular}{l l}
\hline
Node & {Core Score} \\
\hline
King's Cross St.~Pancras & 1.0000  \\

Farringdon & 0.9773 \\

Barbican & 0.9751 \\

Paddington & 0.9693   \\

Great Portland Street & 0.9692  \\

Moorgate & 0.9663  \\

Embankment & 0.9653  \\

Euston Square & 0.9632  \\

Edgware Road & 0.9546 \\

Baker Street & 0.9490  \\
   \hline
\end{tabular}}
\end{center}
\end{minipage}
\caption{The ten most core-like nodes in the London Underground network along with their aggregate core scores (\ref{totalcorescore}), which we obtained using the core quality (\ref{R}) and the transition function (\ref{trans1}).
}\label{tuberesult}
\end{table}
\end{center}

\begin{figure}[!h]
\begin{center}
\begin{minipage}[c]{0.4\linewidth}
\centering
\includegraphics[scale=.55]{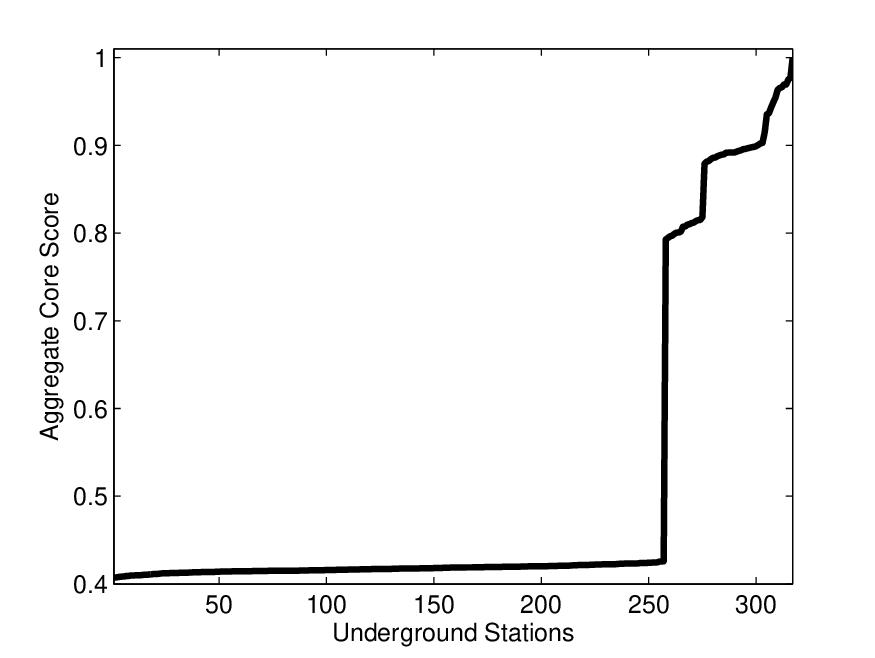}\\(a)
\end{minipage}\hspace*{3pt}
\begin{minipage}[c]{0.4\linewidth}
\centering
\includegraphics[scale=.55]{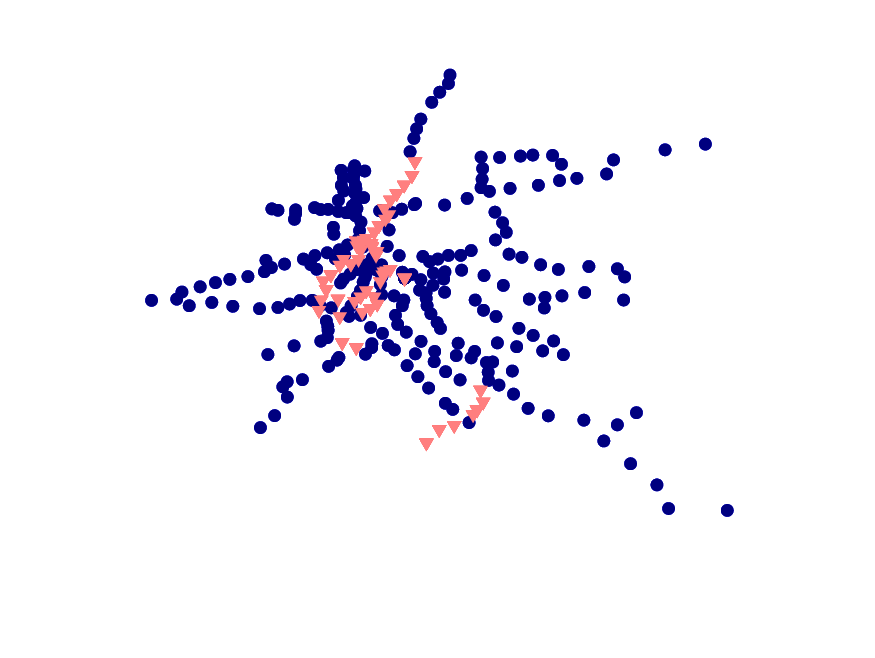}\\(b)
\end{minipage}
\end{center}
\caption{(a) The ordered list of aggregate core scores (\ref{totalcorescore}) for the London Underground stations suggests that there are 60 important stations. [We use the core quality (\ref{R}) and the transition function (\ref{trans1}).] (b) We plot the stations using their geographical locations. The {\color{red}$\filledmedtriangledown $} symbol designates the 60 most important stations, and the {\color{blue}$ \medbullet$} symbol designates the 257 other stations.
} \label{tubepic}
\end{figure}

In Fig.~\ref{tubepic}(a), we plot the aggregate core scores for the stations in order of ascending values. This reveals a sharp jump in aggregate core score and thereby suggests that the London Tube has a core group of (about) 60 stations and a periphery of 257 stations. Additionally, we note that considering core-periphery structure also makes it possible to distinguish between peripheral stations with the same degree centrality. (In the ordering from largest to smallest degree, stations 240--287 all have the same degree.) In Fig.~\ref{tubepic}(b), we plot the stations using their geographical locations. The {\color{red}$\filledmedtriangledown$} symbol designates the 60 most important stations, and the {\color{blue} $\medbullet$} symbol designates the 257 other stations.  In this example, we see that it is reasonable to construe the network as dichotomized into (about) 60 core nodes and (about) 257 peripheral nodes. The large set of {\color{red}$\filledmedtriangledown $} nodes in the middle constitute the stations in Central London (e.g., King's Cross/St. Pancras and Paddington, which are both associated with major train stations). The {\color{red}$\filledmedtriangledown $} nodes that are farther towards the bottom right constitute the stations around Waterloo, which is another major train station in London. A possible explanation for the split core is that the two clusters of core stations are separated geographically by the river Thames, which runs through central London. Most of the historical landmarks (e.g., Buckingham Palace, Trafalgar Square, and the Tower of London) are north of the Thames. The so-called ``South Bank" (which is centered around Waterloo) is a 1960s arts hub containing the Royal Festival Hall, the National Theatre, and the London Eye.

\subsection{Networks of Network Scientists}\label{nns1}

We now consider networks of co-authorships between scholars who study network science. We study two such networks---one from 2006~\cite{New06} and another from 2010~\cite{Rosv10}.  These networks (which both concentrate on papers written by physicists) have $379$ and $552$ nodes, respectively, in their largest connected components. The nodes correspond to scholars working in the field of network science, and an edge between two of them has a weight based on the number of papers that they have co-authored. (Note that the 2006 network is not a subset of the 2010 network.)

In Table \ref{nnscs} of the Appendix, we show the names of the scholars from both 2006 and 2010 with the top thirty aggregate core scores (\ref{totalcorescore}) using the core quality (\ref{R}) and the transition function (\ref{trans1}). In Table \ref{ns2010-newfuncs} in the Appendix, we give the top 30 aggregate core scores for the 2010 network using three variant computations: (left) using the one-parameter transition function (\ref{trans3}) with the product form (\ref{product}) for the core-matrix elements, (middle) using the smooth transition function (\ref{trans2}) with the product form (\ref{product}) and (right) using the usual transition function (\ref{trans1}) with the $p$-norm (\ref{pnorm}) with $p=2$ for the core-matrix elements.  The ordering of the top 30 scholars is similar across different variations of the methodology, although there are some differences.

\begin{figure}[!h]

\begin{center}
\includegraphics[scale=1.0]{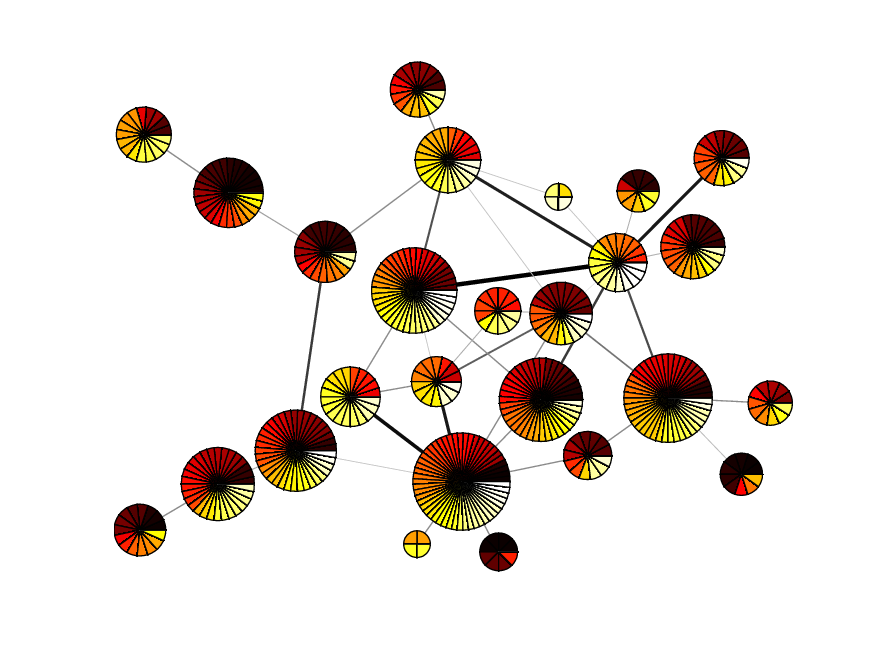}
\end{center}

\caption{Visualization of the 2010 network of network scientists. Each pie represents a community, and the colors represent the rank order of the nodes' aggregate core scores (\ref{totalcorescore}), which we computed using the core quality (\ref{R}) and the transition function (\ref{trans1}).  Darker colors indicate higher rankings; the colors are spaced evenly over all (aggregate) core scores and contain no information about the score distribution.  Each wedge represents a single node, and larger pies contain more nodes. The darkness of the edges represents the total strength of connections between communities. We produced this visualization using code described in Ref.~\cite{Traud09} that uses the Kamada-Kawaii algorithm \cite{Kama88} to locate the centers of the pies. We then tweaked the center locations by hand.}\label{nns}
\end{figure}

The networks of network scientists have both a sensible community structure and a sensible core-periphery structure [recall the block model in Fig.~\ref{comcorim}(c) and (d)].  We illustrate this point in our visualization of the network in Fig.~\ref{nns}. Each pie chart represents a community, which we computed by optimizing modularity using the Louvain algorithm \cite{Blond08}.  Each pie is composed of the nodes in a single community, and each node is represented by a segment colored according to its aggregate core score (\ref{totalcorescore}) computed using the core quality (\ref{R}) and the transition function (\ref{trans1}).  One can plainly see that the network's core nodes are distributed throughout the various communities and that many communities have both core and periphery nodes.

We calculated community structures in which the 2006 network is split into 19 communities and the 2010 network is split into 25 communities, although different community-detection methods yield somewhat different partitions of the networks \cite{good2010}.  For example, one previous examination \cite{richardson09} of community structure in the 2006 network of network scientists using a spectral tripartioning method identified three large groups: one in which A.-L. Barab\'asi is the key node (in the sense of having the largest `community centrality' \cite{New06} in the group), one in which M.~E.~J. Newman is the key node, and one in which A. Vespignani and R. Pastor-Satorras are the two key nodes.  As shown in Table  \ref{nnscs} in the Supplementary Information, all four of these nodes have very high aggregate core scores.

Individual communities in both the 2006 and 2010 networks exhibit a core-periphery structure. As indicated above, the core nodes are distributed throughout the communities. In the 2006 network, 12 of the 19 communities contain at least one node among those with the top 30 aggregate core scores in Table \ref{nnscs}. In the 2010 network, 9 of the 25 communities contain at least one node in the top 30 from Table \ref{nnscs}.  Additionally, each of the communities in the two networks includes one or two highly connected (i.e., high-strength) nodes and several other nodes with low strengths.  In the 2006 network, the mean strength is 4.8, and 17 of the 19 communities contain a node with a strength of at least 9.  (There are 43 such nodes in the entire network.) In the 2010 network, the mean strength is 4.7, and 20 of the 25 communities contain a node with a strength of at least 10. (There are 50 such nodes in the entire network.) This network is an example that contains both an identifiable community structure and an identifiable core-periphery structure.  However, methods to detect core-periphery structure need not indicate anything about community structure and vice-versa. As we discussed previously, community structure and core-periphery structure provide different lenses with which to view a network \cite{jure2013}. There can be examples in which a core and a periphery are describable as separate communities, but community structure and core-periphery structure are different concepts.

In Fig.~\ref{barabasi}, we zoom in on the largest community (53 nodes) in the 2010 network of network scientists.  This community includes the node (A.-L. Barab\'{a}si) with the highest aggregate core score. This figure illustrates that nodes with high scores indeed occupy a well-connected position inside their community as well as in the entire network.

\begin{center}
\begin{figure}[H]
\begin{minipage}[c]{\linewidth}
\begin{center}
\includegraphics[scale=.45]{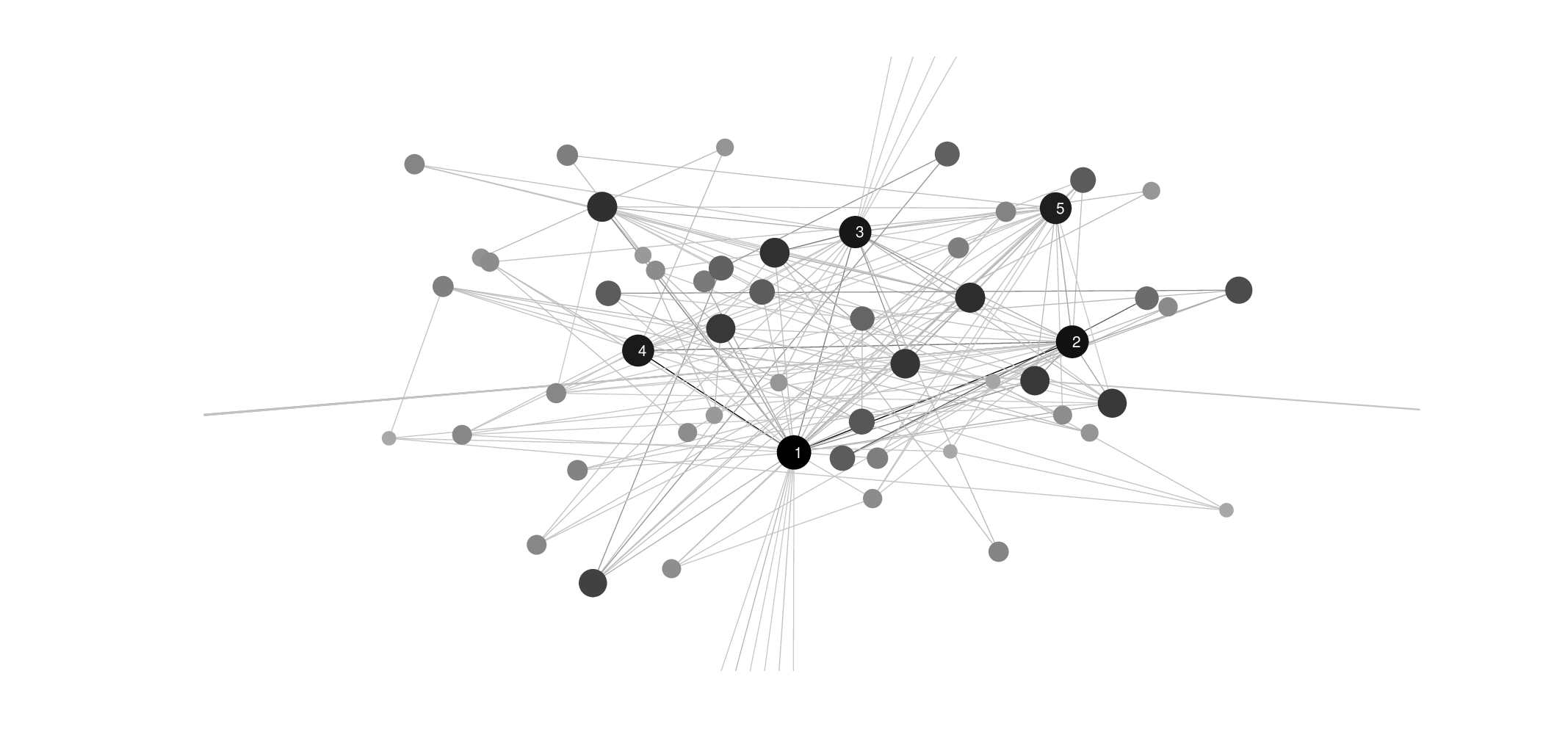}
\end{center}
\end{minipage}
\caption{Magnification of the largest community in the 2010 network of network scientists. The darkness of the edges corresponds to the strength of the edges, and the size and darkness of the nodes represent the aggregate core score.  (Edges that leave the picture are connected to nodes in other communities.) The five labeled nodes, and their corresponding core scores, are A.-L. Barab\'{a}si (1), H. Jeong (.9181), T. Vicsek (.8856), R. Albert (.8737), and Z.~N. Oltvai (.8550).}
\label{barabasi}
\end{figure}
\end{center}

\subsection{Voting-Similarity Network of the United States Senate}\label{voting}

Finally, we consider similarity networks constructed using roll-call votes from the United States Congress.  One can build such a network from a single 2-year Congress of either the Senate or the House of Representatives \cite{voteview,pr97,Waug10}. For each House and Senate, one constructs a complete (or almost complete) weighted network in which each node represents a legislator and a weighted edge between two legislators indicates the similarity of their voting patterns. In our calculation, each adjacency-matrix element $A_{ij}$ is equal to the number of times that legislator $i$ and $j$ voted in the same way divided by the total number of bills on which both $i$ and $j$ cast a vote. This type of network is called a  `similarity network', because the weights of the edges give a measure of similarity between the nodes to which they are incident.  (As was recently discussed in the context of resolutions in the United Nations General Assembly \cite{macon2012}, one can also construct networks from voting data in several other ways.)

As an example, we consider the similarity network for the $108^{\mathrm{th}}$ Senate, which occurred during the third and fourth year of George W. Bush's presidency (2003--2005).  In Table \ref{sen108} of the Appendix, we give for each Senator the aggregate core score (\ref{totalcorescore}) computed using the core quality (\ref{R}) and the transition function (\ref{trans1}). In Fig.~\ref{comparison}, we show scatter plots between the strength centrality and various other centrality measures for the $108^{\mathrm{th}}$ Senate network.  We color Republicans in red and Democrats in blue. The strong similarity between the MINRES and the PageRank computation arises because this example is a similarity network as well as the fact that the aggregate core scores are relatively close together. (See the definition of MINRES in Section \ref{minres}.) They need not be similar in general in other examples.

\begin{figure}[!h]
\begin{center}
\begin{minipage}[c]{0.45\linewidth}
\centering
\includegraphics[scale=.55]{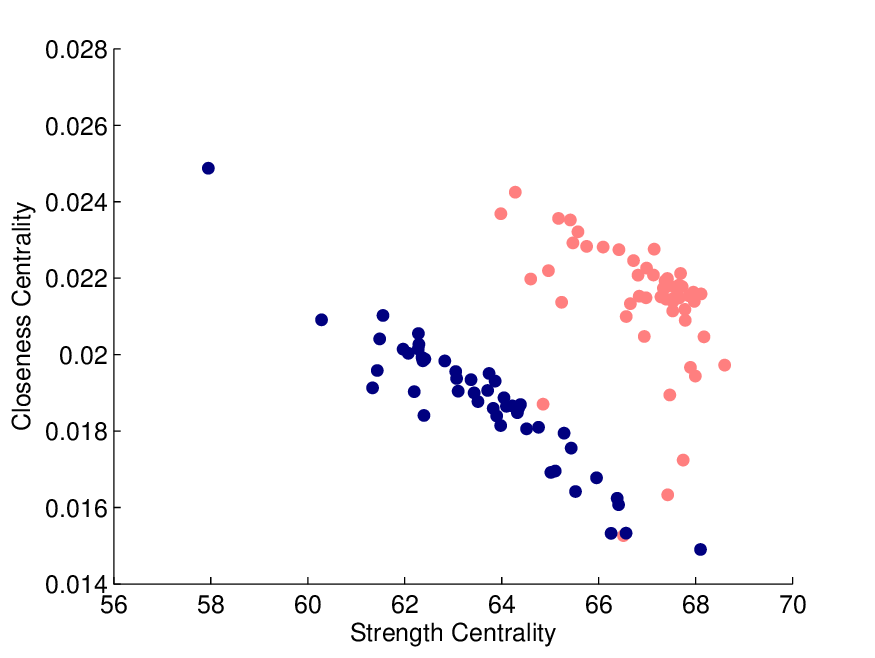}\\(a)
\end{minipage} \hspace*{3pt}
\begin{minipage}[c]{0.45\linewidth}
\centering
\includegraphics[scale=.55]{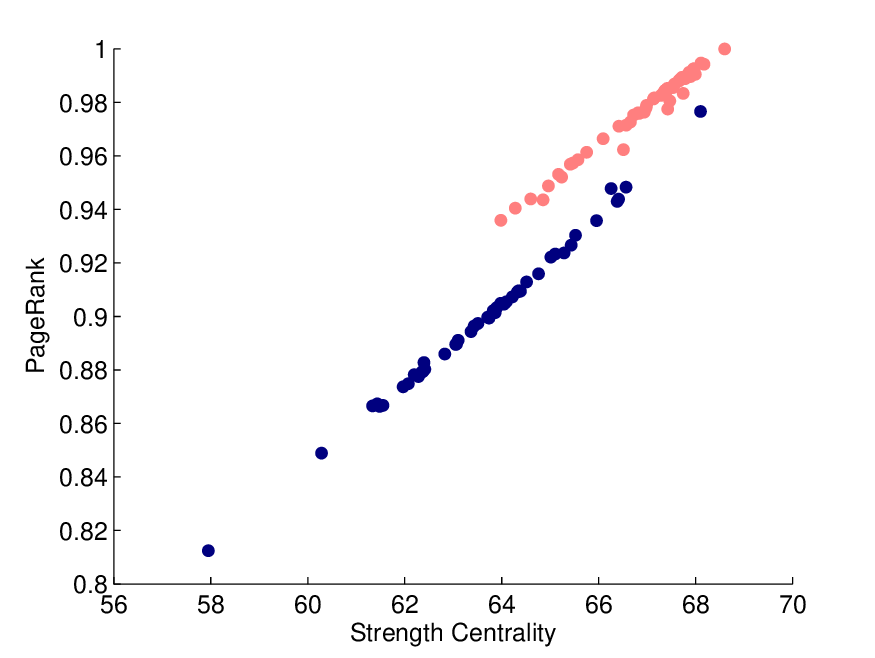}\\(b)
\end{minipage}\hspace*{3pt} \\
\begin{minipage}[c]{0.45\linewidth}
\centering
\includegraphics[scale=.55]{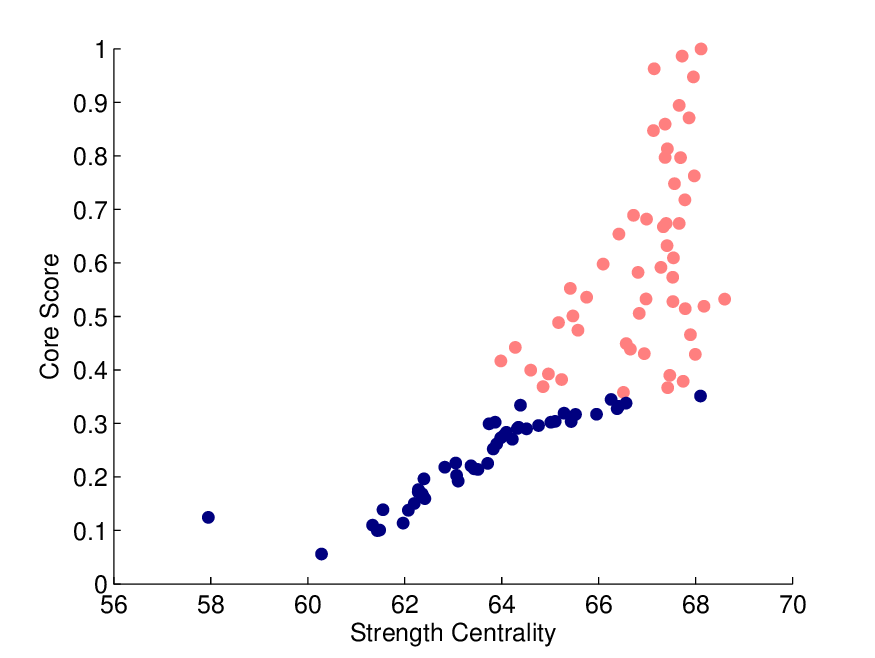}\\(c)
\end{minipage}\hspace*{3pt}
\begin{minipage}[c]{0.45\linewidth}
\centering
\includegraphics[scale=.55]{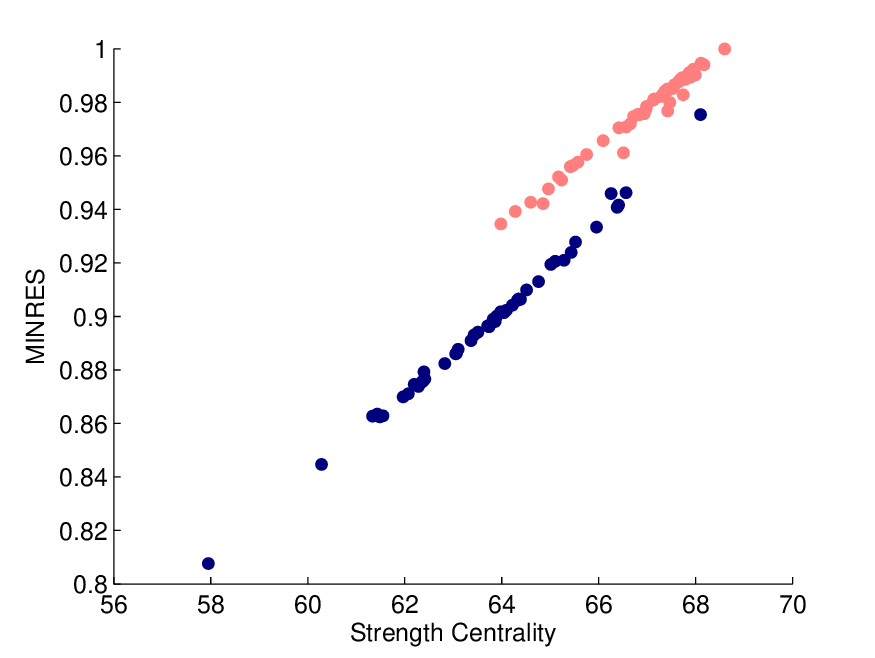}\\(c)
\end{minipage}
\end{center}
\caption{Scatter plots between strength and various other centrality measures for the $108^{\mathrm{th}}$ Senate voting-similarity network.  We show Republicans in red and Democrats in blue. In panel (c), we computed aggregate core scores (\ref{totalcorescore}) using the core quality (\ref{R}) and the transition function (\ref{trans1}). 
}\label{comparison}
\end{figure}

Some of the centrality measures in Fig.~\ref{comparison} have been used previously to study Senators and Representatives in legislation cosponsorship networks \cite{fowler06a,fowler06b}, which have in turn been compared to modularity-based measures of political partisanship studied using roll-call voting networks \cite{yan08}.  As one can see from Fig.~\ref{comparison}, the different centrality measures do indeed measure different things. Observe in particular that none of the centrality measures by themselves separate the communities very well, whereas a combination of two of them can sometimes distinguish a community of (mostly) Republicans and a community of (mostly) Democrats. Investigation of core-periphery structure using aggregate core scores thus complements examination of community structure by allowing one to examine a different type of meso-scale structure. As panel (c) illustrates, it also nicely complements existing centrality measures.

\section{Conclusions and Discussion}\label{concl}

We have proposed a new family of methods to investigate core-periphery structure in networks.  We generalized ideas from Borgatti and Everett \cite{Borg99} and designed an approach that gives nodes values (i.e., core scores) along a continuous spectrum between nodes that lie most deeply in a network core or at the far reaches of a network periphery.  Our approach can be used with a wide variety of different functions to transition between core and peripheral nodes, and it also allows one to use different ways to measure core quality.  The importance of such flexibility, combined with the ability to use our method either to produce a centrality measure for coreness or discrete divisions of core and periphery nodes, has long been recognized by sociologists as an important aspect of core-periphery structure \cite{smithwhite}.

Our investigation of core-periphery structure complements studies of network community structure, which has been considered at great length and from myriad perspectives \cite{Port09,Fort09}.  By contrast, there are comparatively few methods to study core-periphery structure, which we believe is just as important as community structure.  As we have illustrated, networks can contain community structure, core-periphery structure, both, or neither. For example, the 2006 and 2010 networks of network scientists exhibit both types of meso-scale structures in a meaningful way.  In these networks, investigating core-periphery structure reveals a global `infrastructure' that remains invisible if one searches only for community structure.

In contrast to the wealth of attention given to community structure over the last decade, the development of methods to examine core-periphery structure is in its infancy. The purpose of the present paper is conceptual development, and our current implementation of the method is slow because we use simulated annealing.  Additionally, when using two-parameter transition functions, we used 10000 different (and uniformly-spaced) values of $(\alpha,\beta)$, and one can improve speed considerably by considering fewer parameter values, designing schemes to sample values of $\alpha$ and $\beta$ intelligently, or employing a one-parameter transition function.  Further investigations of how to choose core-matrix elements is also important, and one can also investigate core-periphery structure using perspectives that are rather different from the perspective on which we focus in this paper.

Many networks contain meso-scale structures in addition to (or instead of) community structure, and the pursuit of methods to investigate them should prove fruitful.  As we have illustrated, core-periphery structure provides one example that is worth further attention.

\subsection{Acknowledgements}

We thank Alex Arenas, Charlie Brummit, Mihai Cucuringu, Valentin Danchev, Sergey Dorogovtsev, Andrew Elliott, Martin Everett, Des Higham, Sang Hoon Lee, Jos\'{e} Mendes, Jim Moody, Alex Pothen, Stan Wasserman, and two anonymous referees for helpful comments.  We thank Christian Lohse for the suggestion of using the $p$-norm as functional form. We also thank Andrew Elliott for extensive discussions about code. This work was funded by the James S. McDonnell Foundation (\#220020177) and the NSF (DMS-0645369) and was carried out in part at the Statistical and Applied Mathematical Sciences Institute in Research Triangle Park, North Carolina.  We thank Mark Newman for providing the data for the Zachary Karate Club network and the 2006 network of network scientists, Martin Rosvall for providing the data for the 2010 network of network scientists, and Keith Poole and Howard Rosenthal for maintaining the Congressional voting data at {\tt www.voteview.com} \cite{voteview}. The {\sc Matlab} code that we used for simulated annealing was written by Joachim Vandekerckhove~\cite{Kerck}. The {\sc Matlab} code that we used for finding PageRank centrality was written by David Gleich~\cite{Gleich,gleich09}.

\bibliographystyle{siam}
\bibliography{refs7}

\newpage

\section*{Appendix}

\subsection*{Simulated Annealing}

The {\sc Matlab} code that we used for simulated annealing was written by Joachim Vandekerckhove~\cite{Kerck}.  It uses the following parameters: an initial temperature of $1$; a final temperature of $10^{-8}$; a cooling schedule of $.8 \times T$ (where $T$ represents the temperature); a maximum number of consecutive rejections of 1000; a maximum of 300 tries at one given temperature; and a maximum of 20 successes at one given temperature.

\subsection*{Network of Network Scientists}

In Table \ref{nnscs}, we list the names and aggregate core scores (\ref{totalcorescore}) of the top 30 nodes for both the 2006 and 2010 network of network scientists.  To compute the values in this table, we used the core quality (\ref{R}) and the transition function (\ref{trans1}).

In Table \ref{ns2010-newfuncs} in the Appendix, we list the top 30 aggregate core scores for the 2010 network using three variant computations: (left) using the one-parameter transition function (\ref{trans3}) with the product form (\ref{product}) for the core-matrix elements, (middle) using the smooth transition function (\ref{trans2}) with the product form (\ref{product}) and (right) using the usual transition function (\ref{trans1}) with the $p$-norm (\ref{pnorm}) with $p=2$ for the core-matrix elements.   

\begin{center}
\begin{table}[h]

\begin{center}
{\footnotesize

\begin{tabular}{ l l l l }
\hline
NNS2006 Node &{Core Score} & NNS2010 Node &{Core Score}\\
\hline
Barab\'asi, A.-L. &1.00 & Barab\'asi, A.-L. &1.00  \\

 Oltvai, Z. N.&0.97 & Newman, M. E. J. &0.94 \\

 Jeong, H.&  0.96& Pastor-Satorras, R. &0.93 \\

Vicsek, T. & 0.95 & Latora, V. &0.93  \\

Kurths, J. &0.88 & Arenas, A. &0.93\\

Neda, Z. &0.87 & Moreno, Y.	& 0.92\\

Ravasz, E. & 0.86 & Jeong, H. &0.92\\

Newman, M. E. J.&	0.86 & Vespignani, A.	&0.91\\

Pastor-Satorras, R.	& 0.85 & D\'{i}az-Guilera, A.  &0.90\\

Schubert, A. & 0.85 & Guimer\`{a}, R. &0.90\\

Boccaletti, S. 	& 0.85 & Watts, D. J.	&0.89\\

Vespignani, A.	& 0.84 & Vazquez, A. &0.89\\

Farkas, I.	& 0.84 & Viczek, T.& 0.89\\

Derenyi, I.	& 0.83 & Amaral, L. A. N.	&0.89\\

Holme, P.   &	0.82 & Sol\'{e}, R. V. &0.88\\

Crucitti, P.	&0.81 & Albert, R.&	0.87\\

Albert, R.	&0.80 & Kahng, B.&0.87\\

Schnitzler, A.	&0.80 & Boccaletti, S.	&0.86\\

Sol\'{e}, R. 	&0.80 & Oltvai, Z. N.	&0.86\\

Rosenblum, M. 	&0.79 & Barth\'{e}l\'{e}my, M.		&0.85\\

Tomkins, A.  	&0.79 & Kurths, J.&0.84\\

Moreno, Y. 	&0.78 & Fortunato, S.	&0.84\\

Latora, V.		&0.78 & Marchiori, M.	&0.83\\

Rajagopalan, S. &0.78 &  Kertesz, J.	&0.83\\

Raghavan, P.	&0.77 & Caldarelli, G.&0.82\\

Pikovsky, A. &0.76 & Dorogovtsev, S. N. &0.81\\

Kahng, B. &0.75 & Bogu\~{n}\'{a}, M.	&0.80\\

Diazguilera, A. &0.74 & Goh, K. I. &0.80\\

Vazquez, A. &0.74 & Crucitti, P.&0.80\\

Kim, B. &0.74 &Strogatz, S. H.& 0.80 \\
  \hline
\end{tabular}
}\end{center}
\caption{The 30 nodes with the top aggregate core scores (\ref{totalcorescore}) for the (left) 2006 and (right) 2010 networks of network scientists.  We used the core quality (\ref{R}) and the transition function (\ref{trans1}).
}\label{nnscs}
\end{table}
\end{center}

\begin{center}
\begin{table}[h]

\begin{center}
{\footnotesize

\begin{tabular}{ l l| l l| l l}
\hline
NNS2010 Node &{SP\& PN} & NNS2010 Node &{SmF \& PN} & NNS2010 Node &{ShF \& 2N}\\
\hline
Barab\'asi, A.-L. &1.0000 & Barab\'asi, A.-L. &1.0000 & Barab\'asi, A.-L. &1.0000 \\

 Jeong, H.  & .9868 & Moreno, Y. & .9702 & Newman, M. E. J.   & .9954\\

Vespignani, A.  &  .9859 &Vespignani, A. & .9536& Pastor-Satorras, R. & .9932\\

Pastor-Satorras, R.  &.9851 &Jeong, H. & .9361 & Jeong, H. & .9910 \\

Newman, M. E. J.   & .9788 & Newman, M. E. J. & .9176& Vespignani, A. & .9888 \\

Arenas, A. & .9765 & Arenas, A. & .9129 & Moreno, Y. & .9888 \\

Moreno, Y.  & .9762 & Guimer\`{a}, R.  & .8942& D\'{i}az-Guilera, A. & .9862 \\

 Latora, V.  & .9649& D\'{i}az-Guilera, A.& .8809& Latora, V. & .9829 \\

Guimer\`{a}, R.   &  .9638 & Pastor-Satorras, R.& .8755 & Arenas, A. & .9819 \\

Vazquez, A.   & .9616 & Boccaletti, S. & .8686 & Sol\'{e}, R.V. & .9812 \\

D\'{i}az-Guilera, A.  & .9604 & Vicsek, T. & .8355 & Amaral, L. A. N. & .9773 \\

Vicsek, T.  & .9491 & Amaral, L. A. N. & .8341 & Boccaletti, S. & .9768 \\

Amaral, L. A. N.  & .9470 & Latora, V. & .8130 & Vicsek, T. & .9737 \\

Albert, R.  & .9415 & Barth\'{e}l\'{e}my, M. & .8107 & Guimer\`{a}, R. & .9712 \\

 Boccaletti, S.  & .9379 & Vazquez, A. & .8069 & Vazquez, A. & .9689 \\

Watts, D. J.   & .9346 & Kurths, J. & .7714 & Kahng, B. & .9679 \\

Sol\'{e}, R. V. & .9321 & Kahng, B. & .7633 & Kurths, J. & .9676\\

Kahng, B. & .9309 & Oltvai, Z. N.  & .7616 & Kertesz, J. & .9624\\

Kurths, J.  & .9241 & Caldarelli, G. & .7462 & Bornholdt, S. & .9577 \\

Oltvai, Z. N.   & .9197 & Kertesz, J. & .7096 & Dorogovtsev, S. N. & .9554\\

Barth\'{e}l\'{e}my, M. & .9183 & Albert, R. & .7023 & Marchiori, M. & .9549\\

 Marchiori, M. & .9167 & Watts, D. J. & .6861 & Watts, D. J. & .9526 \\

 Caldarelli, G. & .9022 & Porter, M. A. & .6842 & Albert, R. & .9493\\

Kertesz, J.   & .8914 & Sol\'{e}, R. V. & .6823 & Barth\'{e}l\'{e}my, M. & .9488 \\

 Fortunato, S. & .8883 & Fortunato, S. & .6761 & Oltvai, Z. N. & .9478 \\

Goh, K. I.  & .8852 & Kaski, K. & .6752 & Caldarelli, G. & .9474 \\

Kim, D. & .8836 & Tomkins, A. S. & .6648 & Havlin, S. & .9458 \\

Danon, L.  & .8773 & Bogu\~{n}\'{a}, M. & .6584 & Mendes, J.F.F. & .9443\\

 Bogu\~{n}\'{a}, M. & .8747 & Goh, K. I. & .6458 & Stauffer, D. & .9408 \\

Strogatz, S. H.  & .8690 & Kim, D. & .6411 & Tomkins, A. S. & .9401 \\
    \hline
\end{tabular}
}\end{center}
\caption{The 30 nodes with the top aggregate {core scores} from the 2010 networks of network scientists.  From left to right, we computed these scores using the single-parameter transition function (\ref{trans3}) and the product normalization (\ref{product}) (using the parameter values $\alpha = [0.0001:0.0001:1]$ in {\sc Matlab} notation), the smooth two-parameter function (\ref{trans2}) and the product normalization (using the parameter values $\alpha  = \beta = [0.01:0.01:1]$ in {\sc Matlab} notation), and the sharp two-parameter function (\ref{trans1}) and the 2-norm normalization [i.e., (\ref{pnorm}) with $p = 2$] (again using $\alpha  = \beta = [0.01:0.01:1]$ in {\sc Matlab} notation). Note that the second column in Table \ref{nnscs} uses the sharp two-parameter function and the product norm.
}\label{ns2010-newfuncs}
\end{table}
\end{center}

\subsection*{Voting Similarities in the United States Senate}

In Table \ref{sen108}, we show aggregate core scores for Senators in the $108^{\mathrm{th}}$ Congress. We calculated these core scores using the core quality (\ref{R}) and the transition function (\ref{trans1}).

\begin{center}
\begin{table}
\begin{minipage}{0.4\linewidth}
\begin{center}
{\tiny
\begin{tabular}{ l l l }
\hline
{Node} & {{Core Score}} & {Party Vote}\\
\hline
{{\color{red} $\blacksquare$}} 		Chuck Grassley [R - IA] & 1 &97\% \\

{{\color{red} $\blacksquare$}} Thad Cochran [R - MS] & 0.9864 & 98\% \\

{{\color{red} $\blacksquare$}} Mitch McConnell [R - KY] &0.9628 & 98\% \\

{{\color{red} $\blacksquare$}} 	Pete Domenici [R - NM] & 0.9476 & 96\% \\

{{\color{red} $\blacksquare$}} Bill Frist [R - TN] & 0.8943 & 97\% \\

{{\color{red} $\blacksquare$}} 	Pat Roberts [R - KS] & 0.8712 & 97\% \\

{{\color{red} $\blacksquare$}} Conrad Burns [R - MT] & 0.8595 & 96\% \\

 {{\color{red} $\blacksquare$}} Jim Bunning [R - KY] & 0.8472 & 97\% \\

 {{\color{red} $\blacksquare$}}	Saxby Chambliss [R - GA] & 0.8132 & 97\% \\

 {{\color{red} $\blacksquare$}} 	Orrin Hatch [R - UT] & 0.7969 & 97\% \\

{{\color{red} $\blacksquare$}} 		Bob Bennett [R - UT] & 0.7966 & 97\% \\

{{\color{red} $\blacksquare$}} 	Jim Talent [R - MO] & 0.7625 & 97\% \\

{{\color{red} $\blacksquare$}}	Kit Bond [R - MO] & 0.7481 & 96\% \\

{{\color{red} $\blacksquare$}} 		Ted Stevens [R - AK] & 0.7177 & 96\% \\

{{\color{red} $\blacksquare$}} 		John Cornyn [R - TX] & 0.6890 & 96\% \\

{{\color{red} $\blacksquare$}}	Mike Crapo [R - ID] & 0.6819 & 96\% \\

{{\color{red} $\blacksquare$}} 	Liddy Dole [R - NC] & 0.6739 & 96\% \\

{{\color{red} $\blacksquare$}} 	Sam Brownback [R - KS] & 0.6736 & 96\% \\

{{\color{red} $\blacksquare$}} 	Lamar Alexander [R - TN] & 0.6676 & 97\% \\

{{\color{red} $\blacksquare$}} Larry Craig [R - ID] & 0.6540 & 96\% \\

{{\color{red} $\blacksquare$}} 	George Allen [R - VA]&  0.6323 & 96\% \\

{{\color{red} $\blacksquare$}}	Richard Shelby [R - AL] & 0.6094 & 95\% \\

{{\color{red} $\blacksquare$}} 	James Inhofe [R - OK] & 0.5977 & 96\% \\

{{\color{red} $\blacksquare$}}	Richard Lugar [R - IN] & 0.5918 & 96\% \\

{{\color{red} $\blacksquare$}}	Trent Lott [R - MS] & 0.5822 & 95\% \\

{{\color{red} $\blacksquare$}} 	Chuck Hagel [R - NE]& 0.5732 & 95\% \\

{{\color{red} $\blacksquare$}} 	Craig Thomas [R - WY] & 0.5525 & 95\% \\

{{\color{red} $\blacksquare$}}	Wayne Allard [R - CO] & 0.5357 & 95\% \\

{{\color{blue} $\blacksquare$}} 	Zell Miller [D - GA] & 0.5327 & 38\% \\

{{\color{red} $\blacksquare$}} 	Gordon Smith [R - OR] &  0.5324 & 94\% \\

{{\color{red} $\blacksquare$}}	Lisa Murkowski [R - AK] & 0.5277 & 95\% \\

{{\color{red} $\blacksquare$}}	Norm Coleman [R - MN] & 0.5189 & 94\% \\

{{\color{red} $\blacksquare$}}	John Warner [R - VA] & 0.5145 & 94\% \\

{{\color{red} $\blacksquare$}}	Lindsey Graham [R - SC] & 0.5055 & 94\% \\

{{\color{red} $\blacksquare$}}	Jeff Sessions [R - AL] & 0.5009 & 94\% \\

{{\color{red} $\blacksquare$}} 	Mike Enzi [R - WY] & 0.4885 & 94\% \\

{{\color{red} $\blacksquare$}}	Rick Santorum [R - PA] & 0.4741 & 94\% \\

{{\color{red} $\blacksquare$}}	Ben Campbell [R - CO] & 0.4658& 93\% \\

{{\color{red} $\blacksquare$}} 	Peter Fitzgerald [R - IL]& 0.4492 & 93\% \\

{{\color{red} $\blacksquare$}}	Donald Nickles [R - OK] & 0.4420 & 93\% \\

{{\color{red} $\blacksquare$}} 	Kay Bailey Hutchison [R - TX]& 0.4387 & 92\% \\

{{\color{red} $\blacksquare$}} 	George Voinovich [R - OH] &  0.4305& 92\% \\

{{\color{red} $\blacksquare$}}	Mike DeWine [R - OH] & 0.4290& 91\% \\

{{\color{red} $\blacksquare$}} 	Jon Kyl [R - AZ] & 0.4169& 93\% \\

{{\color{red} $\blacksquare$}} 	John Sununu [R - NH]&0.3996 & 91\% \\

{{\color{red} $\blacksquare$}}	John Ensign [R - NV] &0.3923 & 90\% \\

{{\color{red} $\blacksquare$}} 	Arlen Specter [R - PA] & 0.3898& 85\% \\

{{\color{red} $\blacksquare$}} 	Judd Gregg [R - NH]&0.3821 & 90\% \\

{{\color{red} $\blacksquare$}} 	Susan Collins [R - ME] & 0.3789& 84\% \\

{{\color{red} $\blacksquare$}} 	John McCain [R - AZ] & 0.3687& 84\% \\

  \hline
\end{tabular}}
\end{center}
\end{minipage}
\begin{minipage}{0.45\linewidth}
\begin{center}
{\tiny
\begin{tabular}{ l l l }
\hline
{Node} & {{Core Score}} & {Party Vote}\\
\hline

{{\color{red} $\blacksquare$}} 	Olympia Snowe [R - ME] &0.3667 & 82\% \\

{{\color{red} $\blacksquare$}} 	Lincoln Chafee [R - RI] & 0.3580 & 78\% \\

{{\color{blue} $\blacksquare$}} 	Ben Nelson [D - NE] & 0.3512& 72\% \\

{{\color{blue} $\blacksquare$}} 	John Breaux [D - LA] & 0.3448 & 74\% \\

{{\color{blue} $\blacksquare$}} 	Max Baucus [D - MT] & 0.3378& 82\% \\

{{\color{blue} $\blacksquare$}}	Patty Murray [D - WA] &0.3339 & 95\% \\

{{\color{blue} $\blacksquare$}} 	Mary Landrieu [D - LA] &  0.3322& 85\% \\

{{\color{blue} $\blacksquare$}} 	Blanche Lincoln [D - AR] &0.3274 & 87\% \\

{{\color{blue} $\blacksquare$}} 	Tim Johnson [D - SD] &0.3192 & 94\% \\

{{\color{blue} $\blacksquare$}} 	Mark Pryor [D - AR] &  0.3172& 89\% \\

{{\color{blue} $\blacksquare$}}	Evan Bayh [D - IN] &0.3169 & 86\% \\

{{\color{blue} $\blacksquare$}} 	Kent Conrad [D - ND]&  0.3038& 88\% \\

{{\color{blue} $\blacksquare$}} 	Byron Dorgan [D - ND] & 0.3036& 91\% \\

{{\color{blue} $\blacksquare$}} 	Debbie Stabenow [D - MI] & 0.3023& 96\% \\

{{\color{blue} $\blacksquare$}} 	Tom Carper [D - DE] & 0.3021& 86\% \\

{{\color{blue} $\blacksquare$}} 	Barbara Mikulski [D - MD]	& 0.2994& 96\% \\

{{\color{blue} $\blacksquare$}}	Harry Reid [D - NV] & 0.2960 & 93\% \\

{{\color{blue} $\blacksquare$}}	Tom Daschle [D - SD] &0.2927 & 94\% \\

{{\color{blue} $\blacksquare$}} 	Ron Wyden [D - OR] &0.2904 & 93\% \\

{{\color{blue} $\blacksquare$}} 	Bill Nelson [D - FL]& 0.2899& 93\% \\

{{\color{blue} $\blacksquare$}} 	Maria Cantwell [D - WA] & 0.2836& 95\% \\

{{\color{blue} $\blacksquare$}} 	Chuck Schumer [D - NY] &0.2774 & 94\% \\

{{\color{blue} $\blacksquare$}} 	Jeff Bingaman [D - NM] & 0.2732& 92\% \\

{{\color{blue} $\blacksquare$}}	Herb Kohl [D - WI] & 0.2704& 94\% \\

{{\color{blue} $\blacksquare$}} 	Dianne Feinstein [D - CA] & 0.2616& 92\% \\

{{\color{blue} $\blacksquare$}} 	Mark Dayton [D - MN] &  0.2522& 93\% \\

{{\color{blue} $\blacksquare$}} 	Hillary Clinton [D - NY] &0.2261 & 95\% \\

{{\color{blue} $\blacksquare$}} 	Jay Rockefeller [D - WV] &  0.2254& 93\% \\

{{\color{blue} $\blacksquare$}} 	Chris Dodd [D - CT]&0.2209 & 94\% \\

  {{\color{blue} $\blacksquare$}}	Carl Levin [D - MI] & 0.2181& 95\% \\

{{\color{blue} $\blacksquare$}} 	Joseph Lieberman [D - CT] &0.2154 & 93\% \\

{{\color{blue} $\blacksquare$}} 	Joe Biden [D - DE] & 0.2140& 92\% \\

{{\color{blue} $\blacksquare$}} 	Patrick Leahy [D - VT] & 0.2028& 94\% \\

{{\color{blue} $\blacksquare$}} 	James Jeffords [I - VT] & 0.1964& 88\% \\

{{\color{blue} $\blacksquare$}} 	Daniel Inouye [D - HI]& 0.1921& 93\% \\

{{\color{blue} $\blacksquare$}}	Paul Sarbanes [D - MD] & 0.1765& 96\% \\

{{\color{blue} $\blacksquare$}}	Dick Durbin [D - IL] &0.1732 & 95\% \\

{{\color{blue} $\blacksquare$}} 	Barbara Boxer [D - CA] &0.1718 & 95\% \\

{{\color{blue} $\blacksquare$}}	Jon Corzine [D - NJ] &  0.1686& 95\% \\

{{\color{blue} $\blacksquare$}} 	Edward Kennedy [D - MA] &0.1643 & 94\% \\

{{\color{blue} $\blacksquare$}}	Daniel Akaka [D - HI] &0.1593 & 94\% \\

{{\color{blue} $\blacksquare$}}	Russ Feingold [D - WI] &  0.1505& 91\% \\

{{\color{blue} $\blacksquare$}} 	John Edwards [D - NC] &0.1386 & 96\% \\

{{\color{blue} $\blacksquare$}} 	Jack Reed [D - RI] & 0.1378& 95\% \\

{{\color{blue} $\blacksquare$}}	John Kerry [D - MA] & 0.1246 & 98\% \\

{{\color{blue} $\blacksquare$}}	Tom Harkin [D - IA] &0.1138 & 94\% \\

{{\color{blue} $\blacksquare$}} 	Fritz Hollings [D - SC]&0.1097 & 88\% \\

{{\color{blue} $\blacksquare$}} 	Frank Lautenberg [D - NJ] &  0.1006& 94\% \\

{{\color{blue} $\blacksquare$}} Robert Byrd [D - WV]&0.0997 & 90\% \\

{{\color{blue} $\blacksquare$}} 	Bob Graham [D - FL] &   0.0559& 93\% \\

  \hline
\end{tabular}}
\end{center}
\end{minipage}
\caption{Senators in the $108^{\mathrm{th}}$ Congress along with their aggregate core scores (\ref{totalcorescore}) and the percentage of bills for which they voted in line with their political parties.  We determined the core scores using the core quality (\ref{R}) and the transition function (\ref{trans1}).
}\label{sen108}
\end{table}
\end{center}

\end{document}